\documentclass{aa}
\pdfoutput=1

\usepackage{graphicx}
\usepackage{subfig}
\usepackage[varg]{txfonts}
\usepackage{lscape}
\usepackage{color}
\usepackage{longtable}
\graphicspath{{figures/}}

\newcommand{\logg}{\log g}

\def\m2s2{\hbox{\,m$^{2}$\,s$^{-2}$}} 

\begin{document}

   \title{HADES RV Programme with HARPS-N at TNG\thanks{Based on: observations made with the Italian \textit{Telescopio Nazionale Galileo} (TNG), operated on the island of La Palma by the INAF - \textit{Fundaci\'on Galileo Galilei} at the \textit{Roque de Los Muchachos} Observatory of the \textit{Instituto de Astrof\'isica de Canarias} (IAC); photometric observations made with the APACHE array located at the Astronomical Observatory of the Aosta Valley; photometric observations made with the robotic telescope APT2 (within the EXORAP programme) located at Serra La Nave on Mt. Etna.}}
   \subtitle{XV Planetary occurrence rates around early-M dwarfs}
   \titlerunning{Early-M dwarfs occurrence rates}

   \author{M. Pinamonti\inst{\ref{inst1}}
          \and
          A. Sozzetti\inst{\ref{inst1}}
          \and
          J. Maldonado\inst{\ref{inst2}}
          \and
          L. Affer\inst{\ref{inst2}}
          \and
          G. Micela\inst{\ref{inst2}}
          \and
          A. S. Bonomo\inst{\ref{inst1}}
          \and
          A. F. Lanza\inst{\ref{inst3}}
          \and
          M. Perger\inst{\ref{inst4},\ref{inst5}}
          \and
          I. Ribas\inst{\ref{inst4},\ref{inst5}}
          \and
          J. I. Gonz\'alez Hern\'andez\inst{\ref{inst6},\ref{inst7}}
          \and
          A. Bignamini\inst{\ref{inst8}}
          \and
          R. Claudi\inst{\ref{inst9}}
          \and
          E. Covino\inst{\ref{inst10}}
          \and
          M. Damasso\inst{\ref{inst1}}
          \and
          S. Desidera\inst{\ref{inst9}}
          \and
          P. Giacobbe\inst{\ref{inst1}}
          \and
          E. Gonz\'alez-\'Alvarez\inst{\ref{inst11}}
          \and
          E. Herrero\inst{\ref{inst4},\ref{inst5}}
          \and
          G. Leto\inst{\ref{inst3}}
          \and
          A. Maggio\inst{\ref{inst2}}
          \and
          E. Molinari\inst{\ref{inst12}}
          \and
          J. C. Morales\inst{\ref{inst4},\ref{inst5}}
          \and
          I. Pagano\inst{\ref{inst3}}
          \and
          A. Petralia\inst{\ref{inst2}}
          \and
          G. Piotto\inst{\ref{inst13}}
          \and
          E. Poretti\inst{\ref{inst14},\ref{inst15}}
          \and
          R. Rebolo\inst{\ref{inst6},\ref{inst7}}
          \and
          G. Scandariato\inst{\ref{inst3}}
          \and
          A. Su\'arez Mascare\~no\inst{\ref{inst6},\ref{inst7}}
          \and
          B. Toledo-Padr\'on\inst{\ref{inst6},\ref{inst7}}
          \and
          R. Zanmar S\'anchez\inst{\ref{inst3}}
          }

   \institute{INAF - Osservatorio Astrofisico di Torino, Via Osservatorio 20, I-10025 Pino Torinese, Italy\\
              \email{m.pinamonti.astro@gmail.com}\label{inst1}
         \and
             INAF - Osservatorio Astronomico di Palermo, piazza del Parlamento 1, I-90134 Palermo, Italy\label{inst2}
         \and
             INAF - Osservatorio Astrofisico di Catania, Via S. Sofia 78, I-95123 Catania, Italy\label{inst3}
         \and
             Institut de Ci\`encies de l'Espai (ICE, CSIC), Campus UAB, C/ de Can Magrans s/n, E-08193 Bellaterra, Spain\label{inst4}
         \and
             Institut d'Estudis Espacials de Catalunya (IEEC), C/ Gran Capit\`a 2-4, E-08034 Barcelona, Spain\label{inst5}
         \and
             Instituto de Astrof\'isica de Canarias (IAC), E-38205 La Laguna, Tenerife, Spain\label{inst6}
         \and
             Universidad de La Laguna, Dpto. Astrof\'isica, E-38206 La Laguna, Tenerife, Spain\label{inst7}
         \and
             INAF - Osservatorio Astronomico di Trieste, via G. B. Tiepolo 11, I-34143 Trieste, Italy\label{inst8}
         \and
             INAF - Osservatorio Astronomico di Padova, vicolo dell'Osservatorio 5, I-35122 Padova, Italy\label{inst9}
         \and
             INAF - Osservatorio Astronomico di Capodimonte, Salita Moiariello 16, I-80131 Napoli, Italy\label{inst10}
         \and
             Centro de Astrobiolog\'ia (CSIC-INTA), Carretera de Ajalvir km 4, E-28850 Torrej\'on de Ardoz, Madrid, Spain\label{inst11}
         \and
             INAF - Osservatorio Astronomico di Cagliari  \& REM, Via della Scienza, 5, I-09047 Selargius CA, Italy\label{inst12}
         \and
             Dipartimento di Fisica e Astronomia, Universit\`a di Padova, via Marzolo 8, I-35131 Padova, Italy\label{inst13}
         \and
             Fundaci\'on Galileo Galilei - INAF, Ramble Jos\'e Ana Fernandez P\'erez 7, E-38712 Bre\~na Baja, TF, Spain\label{inst14}
         \and
             INAF - Osservatorio Astronomico di Brera, via E. Bianchi 46, I-23807 Merate, Italy\label{inst15}
         }

   \date{Received <date> /
      Accepted <date>}

  \abstract
   {}
   {We present the complete Bayesian statistical analysis of the HArps-n red Dwarf Exoplanet Survey (HADES), which monitored the radial velocities of a large sample of M dwarfs with HARPS-N at TNG, over the last 6 years.}
   {The targets were selected in a narrow range of spectral types from M0 to M3, $0.3$ M$_\odot < M_\star < 0.71$ M$_\odot$, in order to study the planetary population around a well-defined class of host stars. We take advantage of Bayesian statistics to derive an accurate estimate of the detectability function of the survey. Our analysis also includes the application of Gaussian Process approach to take into account stellar activity induced radial velocity variations, and improve the detection limits, around the most-observed and most-active targets. The Markov chain Monte Carlo and Gaussian process technique we apply in this analysis has proven very effective in the study of M-dwarf planetary systems, helping the detection of most of the HADES planets.}
   {From the detectability function we can calculate the occurrence rate of small mass planets around early-M dwarfs, either taking into account only the 11 already published HADES planets or adding also the 5 new planetary candidates discovered in this analysis, and compare them with the previous estimates of planet occurrence around M-dwarf or Solar-type stars: considering only the confirmed planets, we find the highest frequency for low-mass planets ($1$ M$_\oplus < m_p \sin i < 10$ M$_\oplus$) with periods $10$ d$ < P < 100$ d, $f_\text{occ} = 85^{+5}_{-19}\%$, while for short-period planets ($1$ d$ < P < 10$ d) we find a frequency of $f_\text{occ} = 10.3^{+8.4}_{-3.3}\%$, significantly lower than for later-M dwarfs; if instead we take into account also the new candidates, we observe the same general behaviours, but with consistently higher frequencies of low-mass planets. We also present new estimates of the occurrence rates of long-period giant planets and temperate planets inside the Habitable Zone of early-M dwarfs: in particular we find that the frequency of habitable planets could be as low as $\eta_\oplus < 17.1\%$. These results, and their comparison with other surveys focused on different stellar types, confirms the central role that stellar mass plays in the formation and evolution of planetary systems.}
   {}
   

   \maketitle
%

\section{Introduction}
\label{sec:introduction}

Due to the many observational advantages that facilitate the detection of rocky planets in close orbits around them, M dwarfs have become increasingly popular as targets for extrasolar planet search \citep[e.g.][]{dreschar2013,sozzettietal13,astudillodefruetal2017}. These advantages are however counterweighted by the difficulties both in detection and characterisation of exoplanetary systems caused by the stellar activity of the host stars, which can produce radial velocity (RV) signals comparable to those from actual planets, leading to possible misinterpretations \citep[e.g.][]{bonfilsetal2007,robertsonetal2014,anglada16a}. Moreover, M-dwarf planetary systems offer an interesting laboratory to test planet formation theories, since they form under different conditions with respect to FGK-dwarf systems, with different proto-planetary disk masses, temperature and density profiles, and gas-dissipation timescales \citep[e.g.][]{idalin2005}.

Several previous studies were performed aiming to compute the occurrence rate of extrasolar planets of different kinds around low-mass stars, taking into account the detection biases of different surveys. \citet{bonfils13} analysed the M dwarf sample observed as part of the RV search for southern extrasolar planets with the ESO/HARPS spectrograph \citep{mayoretal2003}. \citet{tuomi14} studied a similar, if smaller, sample of M dwarfs observed with both the HARPS and UVES spectrographs \citep{dekkeretal2000}. The exoplanets statistical properties in the Kepler M dwarfs sample were analysed by several authors, e.g. \citet{dressingcharbonneau2015} and \citet{gaidosetal2016}. Their general results pointed towards a high number of low-mass planets orbiting at different distances from their hosts, and confirmed the paucity of giant planets already suggested by earlier surveys \citep{endletal2003,endl06,cumming08}.
Recently, \citet{sabottaetal2021} computed the occurrence rates from a sub-sample of 71 stars observed within the CARMENES exoplanet survey. Their sample covered a wide range of host-star masses, between $0.09$ M$_\odot$ and $0.70$ M $_\odot$, and found slightly larger occurrence rates, although compatible with those from \citet{bonfils13}.

\begin{figure}
\centering
\includegraphics[width=\linewidth]{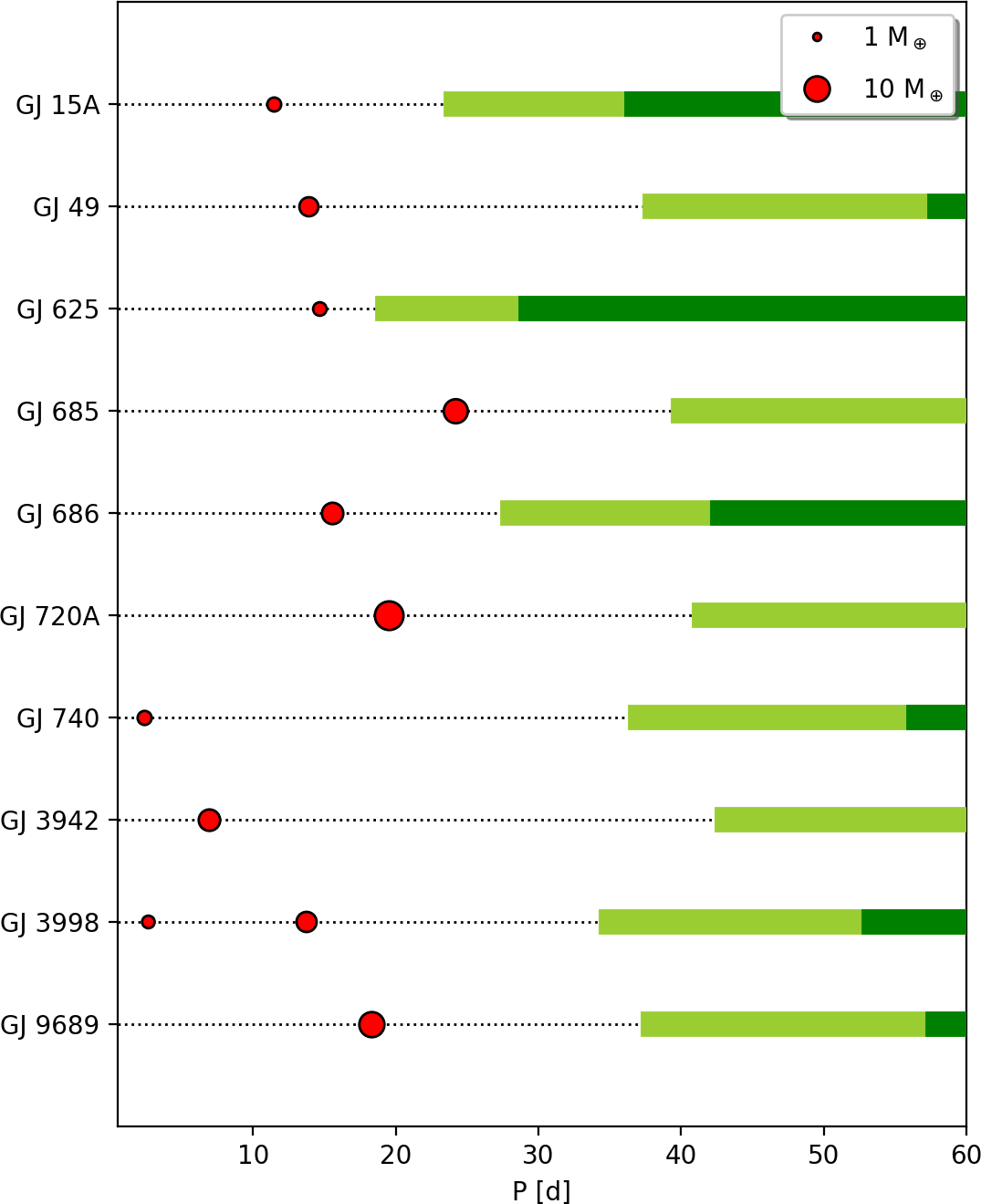}
\caption{Overview of the HADES detected planetary systems. The sample’s published planets are shown as red circles: the symbol size is proportional to the minimum planetary mass. Each system’s Habitable Zone conservative and optimistic limits (see Sect. \ref{sec:hz_occurrence}), are shown as thick dark green and light green bands respectively.}
\label{fig:planet_systems}
\end{figure}

In this framework, the Harps-n red Dwarf Exoplanet Survey \citep[HADES,][]{afferetal2016} programme aims to fully characterize the population of exoplanetary systems on a consistent sample of stars with well known properties, in order to minimize the effects of varying stellar properties on the planetary frequency measurements. To this end, a catalog of early-M dwarfs in narrow range of stellar masses was selected. HADES is a collaboration between the Italian Global Architecture of Planetary Systems \citep[GAPS,][]{covinoetal2013,desideraetal2013,porettietal2016} Consortium, the Institut de Ci\`encies de l'Espai de Catalunya (ICE), and the Instituto de Astrof\'isica de Canarias (IAC). \citet{pergeretal2017} carried out a performance study of the survey, based on the first 4 years of HADES observations: they performed a simulation analysis based on the expected population of early-M planets, and predicted a yield of significant detections of $2.4 \pm 1.5$ planets over the whole sample.
However, the results of \citet{pergeretal2017} were based on the number of observations collected at that time, which were more than doubled by the end of the survey (see Sect. \ref{sec:survey}). Following the same assumptions, in their Fig. 13 they derived the expected yields of planets for different number of targets and observations per target. We can thus derive the expected number of detected planets corresponding to the final number of HADES observations, i.e. $3.8 \pm 1.9$ planets. To date eleven planets have been detected as part of the survey \citep{afferetal2016,suarezmascarenoetal2017,pergeretal2017b,pinamontietal2018,afferetal2019,pergeretal2019,pinamontietal2019,toledopadronetal2021,gonzalezalvarezetal2021,maldonadoetal2021}, and we thus are already exceeding more than 3$\sigma$ from the adjusted prediction. This shows how the previous knowledge of early-M dwarf planetary populations was incomplete. Moreover, during recent years, advanced techniques to mitigate stellar noise have been adopted, further improving the detection limits of RV surveys, thus increasing the yield of detected planets. The published HADES planets are shown in Fig. \ref{fig:planet_systems}. Moreover, in \citet{pinamontietal2019} we studied the population of extrasolar planets orbiting M dwarfs detected through the RV technique. We found moderate evidence of a correlation between planetary mass and stellar metallicity, with different behaviours for small and large planets, which is expected from theory \citep[e.g.][]{mordasinietal2012}. In \citet{maldonadoetal2020} we analysed a large sample of M dwarfs in an homogeneous way, and confirmed that giant planet frequency exhibits a correlation with metallicity, while the same seems not to be true for low-mass planets, these trends are similar to those previously observed for FGK stars \citep[e.g.][]{sousaetal2008,adibekyanetal2012}.  \citet{pinamontietal2019} also confirmed the effect pointed out by \citet{luqueetal2018} of different mass distributions between single and multiple planetary systems. However, these studies were conducted over an heterogeneous sample of planets detected from different surveys, which could not be corrected for detection biases. Therefore their results are to be considered preliminary, since observational biases may have a strong influence on the observed distributions.

In this work we present a thorough Bayesian characterization of the HADES planetary population statistics and global detectability of the survey. We model our analysis on the Bayesian approach proposed by \citet{tuomi14}. We implement this technique with the Monte Carlo Markov chain (MCMC) and Gaussian Process (GP) regression framework, which we successfully applied in the detection of several of the survey's planetary systems \citep[e.g.][]{afferetal2016,pinamontietal2018,pinamontietal2019}.

In Sect. \ref{sec:survey} we describe the HADES programme and summarise the characteristics of its targets. In Sect. \ref{sec:mcmc_analysis} we detail the MCMC-analysis technique and main results in the identification of planetary signals, while in Sect. \ref{sec:detection_limits} we discuss the detection limits of the survey derived from our Bayesian analysis. In Sect. \ref{sec:occurrence_rates} we discuss the planetary occurrence rates we derive for early M dwarfs, and in Sect. \ref{sec:discussion} we compare them with previous results of similar surveys and also with the statistics of FGK-dwarf systems. We summarise our conclusions in Sect. \ref{sec:conclusions}.

\section{The Survey}
\label{sec:survey}

As discussed in \citet{afferetal2016}, the initial HADES sample was constructed by selecting 106 targets from the \citet{lepinegaidos2013} and PMSU \citep[Palomar/Michigan State University,][]{reidetal1995} catalogs: the stars were required to have spectral type between M0 and M3, with masses between $0.3$ M$_\odot$ and $0.7$ M$_\odot$, a visual magnitude $V < 12$, and both to have a high number of Gaia mission \citep{gaiaetal2016} visits and to be part of the target catalog of the photometric survey APACHE \citep[A PAthway towards the CHaracterization of Habitable Earths,][]{sozzettietal13}. Several targets were successively discarded, as they were discovered to be ill-suited for planet search, due to close binary companions, fast rotation ($v\sin i > 4$ km s$^{-1}$), very high chromospheric activity levels ($\log R'_\text{HK} > -4.3$), incorrect spectral type ($T_\text{eff} > 4000$ K), or for having been classified as sub-giants/giants ($\logg < 4$). All the stellar parameters of the targets were consistently re-derived by \citet{maldonadoetal2017} from the HADES HARPS-N measurements applying the techniques from \citet{maldonadoetal2015}, which allowed precise estimates of spectral types, metallicities and effective temperatures. For the current study, we also selected only targets that were observed at least 10 times, since with fewer observations our Bayesian analysis technique discussed in Sect. \ref{sec:mcmc_analysis} could not be robustly applied.

Applying the aforementioned selection criteria,  the HADES sample was reduced to $56$ targets, with spectral type between M0 and M3. The parameters important for this analysis of all the stars in the sample, as computed by \citet{maldonadoetal2020} are listed in Table \ref{tab:star_pam}. The RVs were derived with the Template-Enhanced Radial velocity Re-analysis Application \citep[TERRA,][]{anglada-escudebutler2012}: this template matching approach has proven to be more effective than the CCF technique of the standard HARPS-N Data Reduction System pipeline \citep[DRS,][]{lovispepe2007}, in particular in the analysis of early-M dwarf spectra \citep{afferetal2016,pergeretal2017}. Moreover, we computed the TERRA RVs considering only spectral orders redder than the 22nd ($\lambda$ > 453 nm)\footnote{HARPS-N spectra have a total of 66 échelle orders.}, to avoid the low-SNR blue part of the M dwarfs spectra. The observations were carried out from August 2012 to December 2018, with an average timespan $\left < T_\text{s} \right > = 1760$ d (median $1890$ d). The average number of RVs per target is 77 (median 60), which is more than twice the average number of observations collected at the time of the performance study by \citet{pergeretal2017}. Moreover, it is worth noticing that this typical number of observations per target is significantly larger than the average number of observations used by \citet{bonfils13} (20), and \citet{tuomi14} (62). All RV and activity data used in this analysis are available online in machine readable format.\footnote{All spectroscopic data used in this work will be available at the CDS.}

\begin{table*}
\caption[]{Stellar parameters and summary of the observations of the 56 HADES targets analysed in this work. }
\label{tab:star_pam}
\centering
\begin{tabular}{lcccccccl}
\hline
Star & Sp-Type & M$_\star$ & $[$Fe$/$H$]$ & $N_\text{obs}$ & $T_\text{s}$ & r.m.s. & $P_\text{rot}$ & Reference   \\
& & $[$M$_\odot]$ & & & $[$d$]$ & $[$m$/$s$]$ & $[$d$]$  \\
\hline
  BPM96441  & M$0.0$    & $0.71 \pm 0.09$  & $-0.01 \pm 0.1$  & $17$  & $1295$    & $3.7$ & $18 \pm 4^{*}$    &   \\ 
  TYC3379-1077-1& M$0.0$& $0.69 \pm 0.08$  & $-0.02 \pm 0.09$  & $11 (1)$  & $1393$    & $4.7$ & $23 \pm 4^{*}$    &   \\ 
  GJ 548A   & M$0.0$    & $0.64 \pm 0.08$  & $-0.10 \pm 0.09$  & $35 (3)$  & $909$     & $4.3$ & $36.6 \pm 0.1$    &   \\ 
  StKM1-650 & M$0.0$    & $0.63 \pm 0.08$  & $-0.10 \pm 0.09$  & $20 (1)$  & $2121$    & $24.3$& $15 \pm 3^{*}$    &   \\ 
  GJ 3942   & M$0.0$    & $0.61 \pm 0.07$  & $-0.08 \pm 0.09$  & $145 (5)$ & $1374$    & $6.0$ & $16.3 \pm 0.1$    & {\citet{pergeretal2017b} }\\ 
  NLTT 53166& M$0.0$    & $0.61 \pm 0.07$  & $-0.10 \pm 0.09$  & $31 (1)$  & $1565$    & $3.4$ & $55 \pm 9^{*}$    &   \\ 
  GJ 4057   & M$0.0$    & $0.58 \pm 0.07$  & $-0.19 \pm 0.09$  & $144 (5)$ & $1968$    & $3.2$ & $26.7 \pm 0.1$    &   \\ 
  GJ 731    & M$0.0$    & $0.57 \pm 0.06$  & $-0.16 \pm 0.09$  & $35 (2)$  & $1245$    & $1.9$ & $33 \pm 6^{*}$    &   \\ 
  GJ 3997   & M$0.0$    & $0.49 \pm 0.05$  & $-0.25 \pm 0.09$  & $125 (4)$ & $1896$    & $26.2$& $37 \pm 13$       &   \\ 
  TYC2703-706-1& M$0.5$ & $0.68 \pm 0.06$  & $0.17 \pm 0.09$  & $75 (6)$  & $1593$    & $37.3$& $7.8 \pm 0.2$     &   \\ 
  GJ 4092   & M$0.5$    & $0.66 \pm 0.07$  & $-0.01 \pm 0.09$  & $49 (1)$  & $1921$    & $4.4$ & $25 \pm 5^{*}$    &   \\ 
  GJ 9404   & M$0.5$    & $0.62 \pm 0.07$  & $-0.13 \pm 0.09$  & $54 (3)$  & $1432$    & $4.2$ & $23.2 \pm 0.1$    &   \\ 
  GJ 9689   & M$0.5$    & $0.59 \pm 0.06$  & $-0.10 \pm 0.09$  & $159 (7)$ & $2008$    & $4.6$ & $35.7 \pm 0.1$    &  {\citet{maldonadoetal2021}} \\ 
  GJ 720A   & M$0.5$    & $0.58 \pm 0.06$  & $-0.11 \pm 0.09$  & $118 (3)$ & $1968$    & $4.0$ & $34.5 \pm 4.7$    &  {\citet{gonzalezalvarezetal2021}} \\ 
  GJ 740    & M$0.5$    & $0.58 \pm 0.06$  & $-0.13 \pm 0.09$  & $138 (4)$ & $1968$    & $5.1$ & $36.3 \pm 1.7$    & {\citet{toledopadronetal2021}}  \\ 
  GJ 3822   & M$0.5$    & $0.58 \pm 0.06$  & $-0.12 \pm 0.09$  & $51 (1)$  & $2088$    & $6.9$ & $18.3 \pm 0.1$    &   \\ 
  GJ 3352   & M$0.5$    & $0.57 \pm 0.06$  & $-0.13 \pm 0.09$  & $12$  & $1483$    & $3.7$ & $27 \pm 5^{*}$    &   \\ 
  GJ 685    & M$0.5$    & $0.56 \pm 0.06$  & $-0.12 \pm 0.09$  & $106 (2)$ & $1604$    & $6.2$ & $16.2 \pm 4.2$    & {\citet{pinamontietal2019}} \\ 
  GJ 694.2  & M$0.5$    & $0.56 \pm 0.06$  & $-0.19 \pm 0.09$  & $155 (8)$ & $1599$    & $36.6$& $17.3 \pm 0.1$    &   \\ 
  GJ 21     & M$0.5$    & $0.54 \pm 0.05$  & $-0.11 \pm 0.09$  & $127 (6)$ & $1893$    & $4.5$ & $17.4 \pm 1.1$    &   \\ 
  GJ 1074   & M$0.5$    & $0.54 \pm 0.05$  & $-0.13 \pm 0.09$  & $52 (5)$  & $2071$    & $3.5$ & $25 \pm 5^{*}$    &   \\ 
  GJ 184    & M$0.5$    & $0.53 \pm 0.05$  & $-0.10 \pm 0.09$  & $88$  & $2071$    & $2.7$ & $45.0 \pm	0.1$    &   \\ 
  GJ 162    & M$0.5$    & $0.51 \pm 0.05$  & $-0.19 \pm 0.09$  & $77 (5)$  & $2187$    & $3.4$ & $32.4 \pm 1.6$    &   \\ 
  GJ 412A   & M$0.5$    & $0.38 \pm 0.05$  & $-0.37 \pm 0.09$  & $100 (2)$ & $1558$    & $2.7$ & $100.9 \pm 0.3$   &   \\ 
  GJ 119A   & M$1.0$    & $0.56 \pm 0.05$  & $-0.08 \pm 0.09$  & $129 (5)$ & $1930$    & $3.0$ & $17.4 \pm 1.1$    &   \\ 
  GJ 272    & M$1.0$    & $0.54 \pm 0.05$  & $-0.12 \pm 0.09$  & $13 (1)$  & $2210$    & $5.0$ & $41 \pm 7^{*}$    &   \\ 
  GJ 4306   & M$1.0$    & $0.53 \pm 0.05$  & $-0.14 \pm 0.09$  & $151 (6)$ & $1861$    & $2.8$ & $27 \pm 2.5$      &   \\ 
  GJ 150.1B & M$1.0$    & $0.52 \pm 0.05$  & $-0.12 \pm 0.09$  & $120 (5)$ & $2151$    & $4.7$ & $25 \pm 5^{*}$    &   \\ 
  GJ 3998   & M$1.0$    & $0.52 \pm 0.05$  & $-0.13 \pm 0.09$  & $195 (11)$ & $1570$    & $4.2$ & $33.6 \pm	3.6$    & {\citet{afferetal2016}} \\ 
  GJ 2      & M$1.0$    & $0.50 \pm 0.05$  & $-0.14 \pm 0.09$  & $108 (5)$ & $1864$    & $3.6$ & $21.2 \pm 0.5$    &   \\ 
  GJ 686    & M$1.0$    & $0.41 \pm 0.05$  & $-0.32 \pm 0.09$  & $64 (1)$  & $1347$    & $3.3$ & $70 \pm 12^{*}$   & {\citet{afferetal2019}} \\ 
  GJ 15A    & M$1.0$    & $0.38 \pm 0.05$  & $-0.35 \pm 0.09$  & $123 (5)$ & $1899$    & $2.9$ & $45.0 \pm 4.4$    & {\citet{pinamontietal2018}} \\ 
  GJ 156.1A & M$1.5$    & $0.56 \pm 0.05$  & $-0.03 \pm 0.09$  & $104 (5)$ & $1749$    & $4.2$ & $41.2 \pm 6.3$    &   \\ 
  GJ 49     & M$1.5$    & $0.55 \pm 0.05$ & $-0.03 \pm 0.09$   & $138 (3)$ & $2273$    & $6.2$ & $18.4 \pm 0.7$    & {\citet{pergeretal2019}} \\  
  GJ 9440   & M$1.5$    & $0.52 \pm 0.05$  & $-0.11 \pm 0.09$  & $120 (7)$ & $1594$    & $3.1$ & $48.0 \pm 4.8$    &   \\ 
  GJ 3649   & M$1.5$    & $0.51 \pm 0.05$  & $-0.11 \pm 0.09$  & $25 (2)$  & $2047$    & $3.0$ & $15 \pm 3^{*}$    &   \\ 
  GJ 16     & M$1.5$    & $0.48 \pm 0.05$  & $-0.16 \pm 0.09$  & $125 (5)$ & $2245$    & $3.4$ & $33 \pm 6^{*}$    &   \\ 
  GJ 606    & M$1.5$    & $0.48 \pm 0.05$  & $-0.15 \pm 0.09$  & $29 (3)$  & $1171$    & $4.0$ & $20 \pm 2$        &   \\ 
  GJ 521A   & M$1.5$    & $0.47 \pm 0.05$  & $-0.12 \pm 0.09$  & $144 (7)$ & $2090$    & $2.6$ & $49.5 \pm 3.5$    &   \\ 
  GJ 450    & M$1.5$    & $0.45 \pm 0.05$  & $-0.22 \pm 0.09$  & $40 (2)$  & $1530$    & $3.6$ & $40 \pm 8^{*}$    &   \\ 
  V$^*$BRPsc& M$1.5$    & $0.37 \pm 0.06$  & $-0.29 \pm 0.09$  & $45 (2)$  & $2272$    & $2.6$ & $49.9 \pm 3.5$    &   \\ 
  GJ 1030   & M$2.0$    & $0.51 \pm 0.05$  & $-0.06 \pm 0.09$  & $18 (1)$  & $1960$    & $4.8$ & $32 \pm 3$        &   \\ 
  GJ 414B   & M$2.0$    & $0.51 \pm 0.05$  & $-0.07 \pm 0.09$  & $31 (2)$  & $1039$    & $2.2$ & $62 \pm 10^{*}$   &   \\ 
  NLTT 21156& M$2.0$    & $0.50 \pm 0.05$  & $-0.04 \pm 0.09$  & $44 (3)$  & $2121$    & $17.7$& $10.4 \pm 0.1$    &   \\ 
  GJ 552    & M$2.0$    & $0.47 \pm 0.05$  & $-0.11 \pm 0.09$  & $104 (2)$ & $1530$    & $2.7$ & $43.5 \pm 1.5$    &   \\ 
  GJ 47     & M$2.0$    & $0.36 \pm 0.06$  & $-0.26 \pm 0.09$  & $94 (5)$  & $2280$    & $2.8$ & $34.7 \pm 0.1$    &   \\ 
  GJ 625    & M$2.0$    & $0.30 \pm 0.06$  & $-0.40 \pm 0.09$  & $164 (6)$ & $1921$    & $2.7$ & $77.8 \pm 5.5$    & {\citet{suarezmascarenoetal2017}} \\ 
  GJ 3117A  & M$2.5$    & $0.47 \pm 0.06$  & $-0.06 \pm 0.09$  & $12$  & $1210$    & $2.9$ & $22 \pm 4^{*}$    &   \\ 
  GJ 399    & M$2.5$    & $0.46 \pm 0.06$  & $-0.05 \pm 0.09$  & $37 (2)$  & $2064$    & $3.3$ & $46 \pm 8^{*}$    &   \\ 
  GJ 26     & M$2.5$    & $0.39 \pm 0.06$  & $-0.15 \pm 0.09$  & $55 (2)$  & $2245$    & $3.6$ & $27 \pm 5^{*}$    &   \\ 
  GJ 70     & M$2.5$    & $0.36 \pm 0.07$  & $-0.21 \pm 0.09$  & $27 (2)$  & $1905$    & $2.8$ & $46 \pm 8^{*}$    &   \\ 
  GJ 2128   & M$2.5$    & $0.36 \pm 0.06$  & $-0.26 \pm 0.09$  & $25 (1)$  & $1186$    & $2.4$ & $85 \pm 15^{*}$   &   \\ 
  GJ 408    & M$2.5$    & $0.33 \pm 0.08$  & $-0.21 \pm 0.09$  & $51 (1)$  & $2059$    & $2.2$ & $58 \pm 10^{*}$   &   \\ 
  GJ 119B   & M$3.0$    & $0.45 \pm 0.07$  & $0.03 \pm 0.09$  & $10$  & $1481$    & $1.8$ & $26 \pm 5^{*}$    &   \\ 
  GJ 476    & M$3.0$    & $0.40 \pm 0.07$  & $-0.10 \pm 0.09$  & $16$  & $1300$    & $1.9$ & $55 \pm 5.5$      &   \\ 
  GJ 793    & M$3.0$    & $0.34 \pm 0.08$  & $-0.18 \pm 0.09$  & $27 (2)$  & $1244$    & $2.2$ & $34 \pm 6^{*}$    &   \\ 
\hline
\end{tabular}
\tablefoot{Stellar parameters are from \citet{maldonadoetal2020}. Along with the number of observations $N_\text{obs}$, the numbers in parentheses indicates the numbers of activity-related outliers, if any, identified in each time series as discussed in Sect. \ref{sec:activity_correction}. The listed rotation periods are from \citet{suarezmascarenoetal2017b}; the values derived activity-rotation relationships are marked with a $^{*}$ (see Sect. \ref{sec:rotation}). The last column indicates the references for the published systems.}
\end{table*}

\subsection{Stellar rotation periods}
\label{sec:rotation}

Stellar surface inhomogeneities often produce RV signals close to the rotation period of the star, which can easily mimic the planetary signals \citep[e.g.][]{vanderburg16}. This is particularly significant around M dwarfs, since these stellar signals have periods and amplitudes comparable to those of rocky planets in the habitable zone of their stars \citep{newtonetal2016}. For these reasons, several studies aimed to a precise characterization of the stellar rotation periods and activity features of the HADES sample were performed: \citet{maldonadoetal2017} and \citet{scandariatoetal2017} studied the activity indicators behaviour in the M dwarf sample monitored by the HADES programme, identifying many useful relationships between activity, rotation, and stellar emission lines; \citet{gonzalezalvarezetal2019} studied the X-ray luminosity as a proxy of the coronal activity of the stars in the sample, and its relationship with the stellar rotation periods. \citet{suarezmascarenoetal2017b} instead investigated, using spectral activity indicators and photometry, the presence of signatures of rotation and magnetic cycles in 71 HADES M dwarfs, providing activity-index derived rotation periods for 33 of those stars and 36 rotation periods estimated from activity–rotation relationships. All the 33 targets with activity-derived rotation periods are included in the sample of the current analysis, along with 23 of those with periods derived from activity relationships. The relevant rotation periods are listed in Table \ref{tab:star_pam}.

Moreover, 23 HADES targets have photometric rotation periods detected in the APACHE lightcurves \citep{giacobbeetal2020}, which shows a good correspondence with the spectroscopically-derived rotation periods, being either equal or one close to the first harmonic of the other.

This information over the stellar rotation periods of the sample is very important for our study, since it is paramount to the identification of real Keplerian and activity-induced RV variation during the Bayesian modeling of the time series, as discussed in Sect. \ref{sec:signals}. It is worth noticing that, in the analysis of the impact of activity, the active region lifetime is also an important parameter \citep[e.g.][]{gilesetal2017,damassoetal2019}; however, since not all of our stars have time series well-sampled enough for the application of sophisticated analysis techniques such as GP regression, we focus on the rotation period as the main parameter for stellar activity modeling.

\section{MCMC analysis}
\label{sec:mcmc_analysis}

\subsection{Statistical analysis technique}
\label{sec:stat_analysis}

Our analysis technique follows the approach by \citet{tuomi14}, who used Bayesian statistics to estimate the occurrence rate and detectability function of extrasolar planets around M dwarfs. 

For each time series of the sample, the observed number of planets in a given period-mass interval $\Delta_{P,M}$ can be expressed as:
\begin{equation}
\label{eq:dmsurvey_fobs_eq}
 f_{\text{obs},i}(\Delta_{P,M}) = f_{\text{occ},i}(\Delta_{P,M}) \cdot p_i(\Delta_{P,M}),
\end{equation}
where the subscript $i$ indicates the $i$th system: $f_{\text{occ},i}$ is the occurrence rate of planets around the $i$th star, and $p_i$ is the detectability function, i.e. the probability to detect a planet in the parameter interval $\Delta_{P,M}$ in the $i$th time series. $f_{\text{obs},i}$ is simply computed as the number of planets detected from each time series in each parameter interval.

\begin{figure}
\centering
\includegraphics[width=\linewidth]{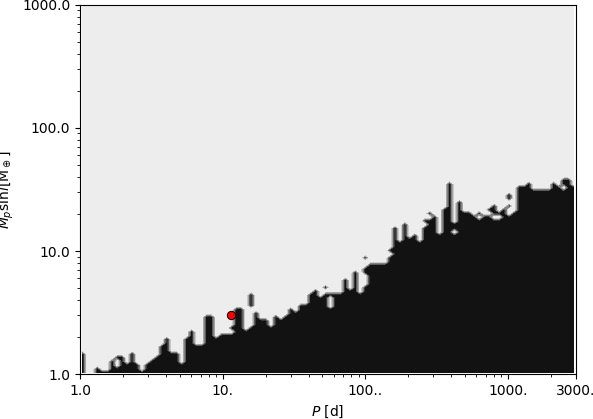}
\caption{Detection function map of the RV time series of GJ 15 A. The white part corresponds to the area in the period - minimum mass space where additional signals could be detected if present in the data, i.e. $p_i = 1$, while the black region corresponds to the area where the detection probability is negligible, i.e. $p_i = 0$. The red circle marks the position in the parameter space of the planet GJ 15 A\,b.}
\label{fig:det_lim_gj15a}
\end{figure}

Assuming that $k_i$ Keplerian signals were detected in the $i$th dataset, the Bayesian technique from \citet{tuomi14} estimates the detectability function $p_i$ by applying a $k_i+1$-signals model to the time series by means of a sampling algorithm. This way, the region of the $(P,M)$ parameter space explored by the parameters of the hypothetical $k+1$th \textit{test-planet}, which cannot be significantly detected in the data, represents the region of the parameter space where the detection technique is not able to reveal significant signals, i.e. the RV amplitudes are so low that the likelihood function does not rule out the presence of an hypothetical additional planetary signal. 
The areas that were not explored by the Markov chains are in fact the areas where signals detection could have been possible, if significant signals were actually present in the data. The detectability function can therefore be approximated as:
\begin{equation}
 p_i = 1-\hat{p}_i,
\end{equation}
where $\hat{p}_i$ is equal to one in the regions of parameters space explored by the $k+1$th Keplerian, and zero otherwise (as detailed in Sect. \ref{sec:test_planet}). An example of the detection function $p_i$, derived with this method, is shown in Fig. \ref{fig:det_lim_gj15a} for the planet-host star GJ 15 A \citep{pinamontietal2018}.

The observed frequency of planets, in a given range of orbital period and planetary mass, over the whole sample of $N$ datasets can be computed by the sum over $i$ of Eq. \ref{eq:dmsurvey_fobs_eq}:
\begin{equation}
\label{eq:dmsurvey_focc_eq}
\begin{split}
 f_\text{obs}(\Delta_{P,M}) & = \sum_{i=1}^{N} f_{\text{obs},i}(\Delta_{P,M}) \\
 & = f_\text{occ}(\Delta_{P,M}) \left [ N - \sum_{i=1}^{N} \hat{p}_i(\Delta_{P,M}) \right],
\end{split}
\end{equation}
assuming the occurrence rate $f_\text{occ}$ to be common for all stars in the sample, $f_\text{occ} = f_{\text{occ},i}$ for all $i$. This is expected to be the case for a well-defined sample of host stars as the HADES sample described in Sect. \ref{sec:survey}.

Since $f_\text{obs} (\Delta_{P,M})$ is known and the square brackets term can be easily calculated, Eq. \ref{eq:dmsurvey_focc_eq} allows to calculate the occurrence rate of planets across the $(P,M)$ parameter space. The global detection probability function of the sample, $p$ can instead be computed simply by dividing the square brackets term by $N$:
\begin{equation}
\label{eq:glob_det_fun}
 p  =  {1 \over N} \left  [ N - \sum_{i=1}^{N} \hat{p}_i(\Delta_{P,M}) \right] = {1 \over N} \sum_{i=1}^{N} p_i(\Delta_{P,M}).
\end{equation}

\subsection{Algorithm implementation}
\label{sec:algorithm}

This Bayesian technique was applied using as sampling algorithm an adapted Python version of the publicly available \texttt{emcee} Affine Invariant MCMC Ensemble sampler by \citet{foreman13}. We also take advantage of Gaussian Process (GP) regression to treat stellar-activity correlated noise from RV time series, which was applied through the GEORGE Python library \citep{ambikasaranetal2015}. These are the same algorithms which have been previously and successfully applied in the study of several HADES targets \citep[e.g.][]{afferetal2016,pinamontietal2018,pinamontietal2019}. We used a variable number of random walkers, $N_\text{walk}$, to sample the parameter space, depending on the number of parameters, $N_\text{par}$, of the final model applied to each time series, always selecting $N_\text{walk} \gtrsim 10 \cdot N_\text{par}$ to ensure a good sampling of the parameter space. 
The convergence of the different MCMC analyses was evaluated calculating the integrated correlation time for each of the parameters, stopping the code after a number of steps equal to 100 times the largest autocorrelation times of all the parameters \citep{foreman13}, after applying an initial burn-in phase \citep{eastman13}.

The general form for the RV models that we applied to all RV time series in this work is given by the equation:

\begin{equation}
\label{eq:general_rv}
\Delta RV(t)  = \Delta RV_\text{BM}(t) + \sum_{j=1}^{N_\text{pl}} \Delta RV_\text{Kep,j}(t) ,
\end{equation}
where $N_\text{pl}$ is the number of planets in the system, the RV Baseline Model, $\Delta RV_\text{BM}(t)$, is defined as:
\begin{equation}
\label{eq:baseline_model}
\Delta RV_\text{BM}(t) = \gamma + d (t-{\bar t}\,) , 
\end{equation}
where $\gamma$ is the systemic velocity, $d$ is the linear acceleration, and $\bar t$ is the mean epoch of the time series. 
The Keplerian RV model $\Delta RV_\text{Kep,j}(t)$ is defined as:
\begin{equation}
\label{eq:rv_kepl}
\Delta RV_\text{Kep,j}(t) =  K_j [\cos(\nu_j(t,e_j,T_{0,j},P_j) + \omega_j) + e_j \cos(\omega_j)] . 
\end{equation}

The log-likelihood function, which is maximized by our fitting algorithm, $\ln \mathcal L$, is defined as follows:

\begin{equation}
\label{eq:rv_likelihood}
\ln \mathcal L = - {1 \over 2} \sum_{t}  \left ( (y(t) - \Delta RV(t)) / \sigma(t)^2 + \ln ( \sigma(t)^2) + \ln(2 \pi) \right ),
\end{equation}
where $y(t)$ is the ``clean'' RV time series (see Sect. \ref{sec:activity_correction}), $\sigma(t)$ is the RV total uncertainty at epoch $t$, computed as $\sigma^{2}(t) = \sigma_\text{data}^{2}(t) +  \sigma_\text{jit}^{2}$, with $\sigma_\text{data}$ the RV HARPS-N internal error, and $\sigma_\text{jit}$ the additional uncorrelated `jitter' term, which is fitted by the model to take into account additional sources of errors such as uncorrelated stellar noise or instrument drifts. For a brief discussion on all the fitted parameters and the adopted priors see Appendix \ref{app:priors}.

\subsubsection{Model selection}
\label{sec:model_selection}

We identified the best model for each time series by computing the Bayesian information criterion \citep[BIC,][]{schwarz1978}, defined as:
\begin{equation}
\label{eq:bic}
\text{BIC} = N_\text{par} \ln(N_\text{obs}) - 2 \ln \mathcal L ,
\end{equation}
and, starting from the RV Baseline Model, $\Delta RV_\text{BM}(t)$, accepting a model with an higher number of parameters when its BIC value was at least 10 points lower than the previous model, which denotes a strong statistical evidence in favour of the lower-BIC model \citep{kassraftery1995}.\footnote{The MCMC-derived detectability function we adopt in this analysis shows good correspondence with the $\Delta \text{BIC} = 10$ detection threshold, as shown in Appendix \ref{app:priors}.} 
 
\subsubsection{Test planet model}
\label{sec:test_planet}

When the best model, containing $N_\text{pl}$ planetary signals, has been identified for each time series, in order to compute the detectability function $p_i$, we run one additional MCMC fit adding one additional `Keplerian' term to Eq. \ref{eq:general_rv}, which corresponds to the \textit{test-planet} used to sample $\hat{p}_i$. The RV model thus becomes:

\begin{equation}
\label{eq:general_rv_test}
\Delta RV(t)  = \Delta RV_\text{BM}(t) + \sum_{j=1}^{N_\text{pl}+1} \Delta RV_\text{Kep,j}(t) ,
\end{equation}
where $\Delta RV_{\text{Kep,}N_\text{pl}+1}(t) = \Delta RV_\text{test}(t)$ is the \textit{test-planet} RV model. It is worth noticing that, since the  $N_\text{pl}$-planet model was already the `best-model' for the corresponding time series, the $N_\text{pl}+1$ test model will not be statistically favoured with respect to the former, i.e. $\Delta \text{BIC} < 10$. Moreover, for the sake of computational efficiency, we restricted the \textit{test-planet} RV model to the circular case, imposing $e_\text{test} = \omega_\text{test} = 0$, i.e.:
\begin{equation}
\Delta RV_\text{test}(t) =  K_\text{test} \cos(\nu_\text{test}(t,T_{0,\text{test}},P_\text{test}))  \notag .
\end{equation}
This assumption should not strongly affect our results, since eccentricity up to 0.5 is not expected to have a strong impact on detection limits \citep{endl02,cummingdragomir2010}.

Finally, from the posterior distribution of $P_\text{test}$, and that of  $K_\text{test}$ converted to minimum-mass, it is possible to compute $\hat{p}_i (P,M)$ as discussed above.

\subsection{Stellar noise correction}
\label{sec:activity_correction}

As previously discussed, an important issue in the analysis of M-dwarf RV time series is the presence of stellar noise. It is known that not all stellar RV variations are directly related to the stellar rotation period \citep[e.g.][and references therein]{dumusqueetal2015}, and thus we had to adopt a multi-step approach to correct as much as possible of the correlated and un-correlated stellar noise in order to improve the estimated planetary parameters, and the derived detectability function.

For all the HADES targets selected for this study, we computed the stellar activity indexes based on the Ca~{\sc ii}  H and K, H$\alpha$, Na~{\sc i} D$_{\rm 1}$ D$_{\rm 2}$, and He~{\sc i} D$_{\rm 3}$ spectral lines, applying the procedure described in \citet{gomesdasilva11} to all the available HARPS-N spectra. We then applied the following two correction criteria to all time series:

\begin{itemize}
 \item outliers removal: first we used the activity index time series to identify any potential outlier caused by transient stellar effect such as flares: this can be very important, since flares have been observed to be very common even around early-to-mid M dwarfs \citep{hawleyetal2014}, and they have been observed to manifest even around stars showing little to no periodic stellar signals produced by active regions \citep{yangetal2017}. To identify these outliers around each target we performed a $3\sigma$ clipping around the median of each activity time series; if any long-term trend was present in the activity time series, the clipping was performed over the linearly de-trended time series, to avoid that the long term variation could enlarge the standard deviation of the data hiding significant outliers. We then removed the epochs corresponding to the identified outliers from all the activity and RV time series for that target. The number of outliers identified in each time series is listed in Table \ref{tab:star_pam}.
 \item correlation with activity indexes: other known stellar phenomena to take into account are magnetic cycles, which have been observed to produce significant variations in RV time series \citep[e.g.][]{lovisetal2011,dumusqueetal2011}. This effect can be identified by the presence of a long-term correlation between the RV and spectroscopic activity indexes time series. For this reason, after the outliers removal, for each target we computed the Pearson correlation coefficients between the RVs and all activity indicators, and if any strong correlation was found ($\rho > 0.5$) we detrended the RV time series via linear-fit with the corresponding index time series.\footnote{If a significant correlation was found for more than one activity index, we selected for the linear detrending the time series which presented the strongest correlation.} We choose to take into account only strong correlations, since even if lower correlations could still be hint of a magnetic cycle contamination in the RVs, they would be more difficult to correct via a simple linear fit, with the risk of injecting additional noise in the time series if the RVs were inappropriately detrended. The time series that were corrected for significant activity correlation, and the correlated activity indexes are listed in Table \ref{tab:best_fit_models}.
\end{itemize}

These corrections were applied to all the analysed systems, producing the ``clean'' RV time series on which the MCMC algorithm was applied to estimate the planetary parameters and compute the detection limits as previously discussed.
In addition to these, we adopted additional steps for those systems with periodic correlated stellar signals in the RV time series corresponding to either the star's rotation period or its first harmonics. We used two alternative techniques to model these stellar signals. The first approach was to fit the activity RV signals as sine-wave, i.e. to include them in the general model of Eq. \ref{eq:general_rv} (in which case the summation is computed over $N_\text{signals} = N_\text{pl} + N_\text{act}$, including the number of activity signals, $N_\text{act}$).

The second approach was to apply GP regression to those systems which presented strong activity RV signals. We adopted the GP regression only for those systems which had a sufficiently large number of observations, in order to avoid overfitting the data due to the adaptive nature of the GP. We chose this threshold to $N_\text{obs} > 70$ RVs per target, in order to have at least $10 \times$ the number of minimum parameters in the model (3 BM $+4$ GP). For those systems, we computed the likelihood function to be maximized by the MCMC sampler via GP regression, i.e.  the log-likelihood function in Eq. \ref{eq:rv_likelihood} was re-defined as:

\begin{equation}
\label{eq:gp_likelihood}
\ln \mathcal L = - {1 \over 2} \mathbf{r}^T \times \mathbf{K}^{-1} \times \mathbf{r} -  {1 \over 2} \ln(\text{det} \mathbf{K}) - {N_\text{obs} \over 2} \ln(2 \pi) ,
\end{equation}
where $\mathbf{r}$ is the residual vector $\mathbf{r} = y(t) - \Delta RV (t)$, with $\Delta RV (t)$ either from Eq. \ref{eq:general_rv} or Eq. \ref{eq:general_rv_test}, and $\mathbf{K}$ is the covariance matrix.
The covariance matrix is defined by the GP kernel: in this work we adopted the commonly used Quasi-Periodic (QP) kernel, defined as the multiplication of an exp-sin-squared kernel with a squared-exponential kernel \citep{ambikasaranetal2015}. The QP kernel can be expressed as follows:

\begin{equation}
\label{eq:gp_kernel}
K_{i,j} = h^2 \cdot \exp \bigg[ -{(t_i-t_j)^2 \over 2 \lambda^2} - {\sin^{2}(\pi(t_i-t_j)/\theta) \over 2 w^2}\bigg] + \sigma^{2} \cdot \delta_{i,j} ,
\end{equation}
where $K_{i,j}$ is the $i j$ element of the covariance matrix, and the covariance is described by four hyper-parameters: $h$ is the amplitude of the correlation, $\lambda$ is the timescale of decay of the exponential correlation, $\theta$ is the period of the periodic component, and $w$ is the weight of the periodic component. The last term of Eq. \ref{eq:gp_kernel} describes the white noise component of the covariance matrix, where $\delta_{i,j}$ is the Kronecker delta and $\sigma$ is the RV total uncertainty defined as in Eq. \ref{eq:rv_likelihood}.

As a final remark, it is important to point out that both these approaches to the fitting of correlated stellar signals were applied only when the model including the stellar-activity correction was preferred over the previous model by our model selection criterion discussed in Sect. \ref{sec:model_selection}. The adopted priors for the GP hyper-parameters are discussed in Appendix \ref{app:priors}, while in Table \ref{tab:best_fit_models} all the time series on which GP regression was applied are listed, as well as all time series in which sinusoidal activity signal were fitted. 

\subsection{Planetary candidates and activity signals}
\label{sec:signals}

We now briefly discuss the Keplerian RV signals and stellar activity features identified in the best-fit models of our analysis. A summary of all the adopted models for each system in the survey is given in the Appendix in Table \ref{tab:best_fit_models}.

\paragraph{Published systems}
We re-analysed with the algorithm discussed in the previous Sections all the published HADES systems (see Fig. \ref{fig:planet_systems}), in order to compute the detection function $p_i$. These independent analyses confirmed the previous results, producing a good agreement in the fitted planetary and activity parameters, usually consistent within $1\sigma$ with the published best-fit values. Some systems required additional care in the re-analysis, since in this work we used only the HADES HARPS-N RV time series collected during the programme, while some publication took advantage of additional RV data from the literature or from other instruments \citep[e.g.][]{pinamontietal2018,pergeretal2019}. In particular in the case of GJ 15 A it was not possible to model the long-period planet GJ 15 A\,c with only the HARPS-N data, which have a much shorter timespan than its orbital period \citep{pinamontietal2018}: we thus decided not include the long-period Keplerian of planet $c$ in the RV model, but to use the acceleration term $d$ of the Baseline Model to fit its contribution in the HADES dataset\footnote{This is a good approximation over the HARPS-N time series, as can be seen in Fig. 6 of \citet{pinamontietal2018}.}.

\paragraph{Planetary candidates}
\label{sec:candidates}
In addition to the already published HADES planetary systems, there are a few other systems in which our analyses identified promising periodic signals, with no obvious relationship with the stellar rotation period or other activity-driven phenomena, and thus could be considered planetary candidates. Since the in-depth analysis required to ascertain their true planetary nature is beyond the scope of this paper, we will not present detailed analyses or complete orbital parameters for these systems, which will be discussed in future specific publications, two of which are currently in preparation: Gonz\'alez Hern\'andez et al. and Affer et al. will discuss GJ 21 and GJ 3822 systems, respectively. However, since these additional planets would affect the occurrence rates derived from our survey, we will take into account the presence of these additional candidates, as discussed in Sect. \ref{sec:occurrence_rates}. The best-fit orbital periods and minimum masses of these candidates are listed in Table \ref{tab:candidates}, and in Appendix \ref{app:candidates_plan} a brief overwiew of the corresponding RV signals is presented. 

\begin{table}
\caption[]{Candidate planetary signals detected in the RV modeling.}
\label{tab:candidates}
\centering
\begin{tabular}{lccp{0.15\textwidth}}
\hline
Target & $P$ & $m_p \sin i$ & Notes \\
 & $[$d$]$ & $[$M$_\oplus]$ \\
\hline
\noalign{\smallskip}
GJ 21 & $30.3^{+0.1}_{-2.4}$ & $5.8^{+1.4}_{-1.4}$ & {\footnotesize Gonz\'alez Hern\'andez et al. (in prep.)} \\    
\noalign{\smallskip}
GJ 1074 & $7.140^{+0.078}_{-0.005}$  & $5.1^{+1.2}_{-1.2}$  \\    
\noalign{\smallskip}
GJ 9404 & $13.46^{+0.01}_{-0.51}$ & $10.3^{+1.8}_{-1.8}$ \\   
\noalign{\smallskip}
GJ 548A & $13.080^{+0.067}_{-0.033}$ & $9.5^{+2.2}_{-2.0}$ \\   
\noalign{\smallskip}
GJ 3822 & $25.151^{+0.081}_{-0.055}$ & $20.9^{+3.1}_{-3.5}$ &  {\footnotesize Affer et al. (in prep.)} \\    
\noalign{\smallskip}
\hline
\end{tabular}
\end{table}

\paragraph{Activity signals}
Many additional significant signals have been identified in the modeling of the remaining HADES time series, which either had a direct counterpart in the activity indexes or were classified as of stellar origin after further investigations. All these activity signals were still included in our models, either as quasi-periodic signals in the GP regression (Eq. \ref{eq:gp_kernel}) or as sinusoidal signals included in the general model in Eq. \ref{eq:general_rv}. A general discussion of the RV signals of stellar origins identified in the HADES time series can be found in \citet{suarezmascarenoetal2017b}.

\subsection{Long-term trends}
\label{sec:trends}

In addition to the periodic RV signal identified during the MCMC analysis, another interesting aspect is the presence of long-term trends in the RV time series: these could indicate the presence of long-period companions, which could not otherwise be detected due to the limited temporal baseline of the survey. To account for this, our RV Baseline Model applied to all systems, $\Delta RV_\text{BM}(t)$, always includes an acceleration term, $d (t-{\bar t}\,)$ (see Eq. \ref{eq:baseline_model}).

In Table \ref{tab:trends} the targets that presented significant ($>3\sigma$) accelerations in the final RV model are listed. It is worth noticing that, as discussed in Sect. \ref{sec:activity_correction}, the analysed RV time series were checked and corrected for correlations with the activity indexes, to reduce long-term signals due to stellar activity or magnetic cycles. However, two of the targets that showed significant long-term trends, GJ 119A and GJ 694.2, were previously identified as having long-term activity cycles, expected to produce the RV variation \citep{suarezmascarenoetal2017b}, even though we did not find significant correlation between RVs and activity indexes. Nevertheless, the RV trend in GJ 694.2's time series has a peak-to-peak amplitude of more than $>100$ m s$^{-1}$, which is extremely large compared to other M dwarfs with detected RV activity cycles, and it is thus probably produced by other sources (see below). 

\begin{table}
\caption[]{Significant linear trends identified in the RV modeling.}
\label{tab:trends}
\centering
\begin{tabular}{lcl}
\hline
Target & $d$ & Notes \\
 & $[$m s$^{-1}$ d$^{-1}]$ & \\
\hline
\noalign{\smallskip}
GJ 16 & $0.00315^{+0.00072}_{-0.00077}$ \\   
\noalign{\smallskip}
GJ 15A & $0.00159^{+0.00049}_{-0.00049}$ & Planet GJ 15A\,c\\
\noalign{\smallskip}
V$^*$BRPsc & $0.00247^{+0.00044}_{-0.00045}$ & \\
\noalign{\smallskip}
NLTT 21156 & $0.0145^{+0.0046}_{-0.0048}$ & \\
\noalign{\smallskip}
\textbf{StKM1-650} & $0.0287^{+0.0008}_{-0.0013}$ & Brown dwarf\\
\noalign{\smallskip}
GJ 793 & $0.0038^{+0.0011}_{-0.0011}$ & \\
\noalign{\smallskip}
GJ 412A & $0.00304^{+0.00046}_{-0.00046}$ & \\
\noalign{\smallskip}
\textbf{GJ 694.2} & $0.0697^{+0.0025}_{-0.0025}$ & Binary \\
\noalign{\smallskip}
GJ 408 & $0.00248^{+0.00046}_{-0.00047}$ & \\
\noalign{\smallskip}
GJ 4092 & $0.00413^{+0.00084}_{-0.00086}$ & \\
\noalign{\smallskip}
\textbf{GJ 3997} & $-0.0323^{+0.0057}_{-0.0057}$ & Brown dwarf \\
\noalign{\smallskip}
\hline
\end{tabular}
\tablefoot{Trends that are not counted as potential long-period planets are highlighted in bold.}
\end{table}

A few systems showed additional long-term evolution, with significant curvature that could not be fitted via a simple linear trend, and thus we introduced in their analyses a quadratic acceleration term, changing the Baseline Model in Eq. \ref{eq:baseline_model} to:
\begin{equation}
\label{eq:baseline_model_quad}
 \Delta RV_\text{BMQ}(t) = \gamma + d (t-{\bar t}\,) + {1 \over 2} q (t-{\bar t}\,)^2,
\end{equation}
where $q$ is the quadratic acceleration coefficient. These quadratic trends were included in the final RV models whenever their addition to the model was accepted by our model selection criteria (see Sect \ref{sec:model_selection}). The significant quadratic long-term trends detected in our sample are listed in Table \ref{tab:parab}. Of these three targets, GJ 119A shows a significant quadratic accelerations, but it is suspected to be related activity cycles previously identified by \citet{suarezmascarenoetal2017b}. Moreover, GJ 649.2 and GJ 3997 have both massive companions which are most probably the source of the observed trends (see below).

\begin{table}
\caption[]{Significant quadratic trends identified in the RV modeling.}
\label{tab:parab}
\centering
\begin{tabular}{lcl}
\hline
Target & $d$ & Notes \\
 & $[$m s$^{-1}$ d$^{-1}]$ & \\
\hline
\noalign{\smallskip}
\textbf{GJ 119A} & $-11.2^{+2.2}_{-2.3} \cdot 10^{-6}$ &  Activity? \\
\noalign{\smallskip}
\textbf{GJ 694.2} & $-19.2^{+3.1}_{-3.1} \cdot 10^{-6}$ & Binary \\
\noalign{\smallskip}
\textbf{GJ 3997} & $-23.0^{+6.1}_{-6.1} \cdot 10^{-6}$ & Brown dwarf \\
\noalign{\smallskip}
\hline
\end{tabular}
\tablefoot{Trends that are not counted as potential long-period planets are highlighted in bold.}
\end{table}

We checked the targets showing long-term trends to identify any stellar companions that could be the origin of the observed RV shift. GJ 793 and GJ 4092 are both single stars with no known stellar companion. GJ119A, GJ412 A, and StKM1-650 have all common proper motion companions, but with projected separations of 340 AU, 160 AU, and 980 AU respectively, the maximum RV acceleration produced by the stellar companions\footnote{The RV acceleration produced by the stellar binary is computed assuming a circular orbit edge-on and using the measured separation as semi-major axis.} is on average one order of magnitude lower than the best-fit $d$ values listed in Table \ref{tab:trends}. Thus the long-term trends observed in these systems cannot be ascribed to the nearby stellar companions, but could in fact be due to the presence of as-of-yet unkown perturbers.
Moreover, GJ 3997 and NLTT 21156 both have another stellar object with a small on-sky separation, but the line-of-sight separation measured from the Gaia EDR3 parallaxes is in both cases $> 10^4$ AU, which again means that the nearby objects could not produce the observed RV trends.

Finally, we checked for Proper Motion Anomalies (PMA) between Gaia EDR3 and Hipparcos \citep{kervellaetal2021}, and other indications of unresolved massive companions in Gaia EDR3, such as RUWE and excess noise \citep{gaiaetal2021}. Three targets show significant PMA, GJ 15A, GJ 694.2, and GJ 119 A: in GJ 15A the anomaly is probabily due to the stellar companion GJ 15B; in GJ694.2 the PMA is caused by a subarcsec stellar companion, which is also the probable source of the measured RV long-term variation; in GJ 119A the PMA is also probably caused by the stellar companion, which is nevertheless too distant to produce the observed RV accelaration (as discussed above). Two of the other  targets, StKM1-650 and GJ 3997, while not having PMA measurements, present high RUWE and excess noise values in Gaia EDR3, which are potential evidence of orbital motion, indicating the presence of possible brown dwarf companions \citep{lindegrenetal2021}.
It is worth noticing that these analyses on the presence of long-period planetary companions in the HADES sample are still preliminary, but a more in-depth study to ascertain the nature of the observed long-term RV trends goes beyond the scope of this paper.

\section{Detection Limits}
\label{sec:detection_limits}

\begin{figure}
\centering
\includegraphics[width=\linewidth]{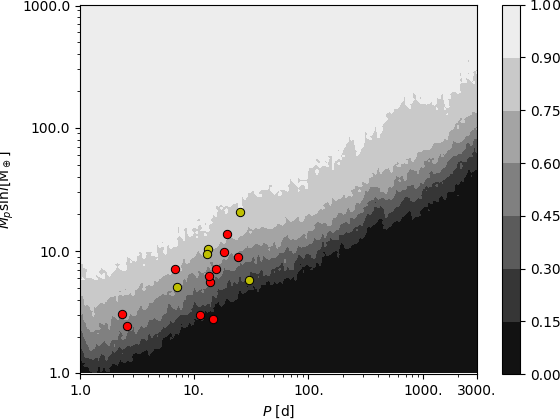}
\caption{Global HADES detection map. The color scale expresses the global detection function, $p$. The red circles mark the position in the parameter space of the  confirmed HADES planets (see Fig. \ref{fig:planet_systems}), while the yellow circles indicate the additional candidates presented in this study (see Table \ref{tab:candidates}).}
\label{fig:det_lim_survey}
\end{figure}

To derive the detectability function, we first mapped the parameter space into a $150 \times 150$ logarithmic grid: the period range covered was $[1,3000]$ d, obtained from the prior of $P_\text{test}$ defined in Appendix \ref{app:priors}; the investigated minimum-mass range was instead $[1,1000]$ M$_\oplus$. We then used this $(P,M)$ grid and the posterior distributions derived from the MCMC analysis of each HADES system to compute $\hat{p}_i$, from which the global detection probability function $p$ of the HADES sample could be computed as the mean of all the systems detection functions, as shown in Eq. \ref{eq:glob_det_fun}. The resulting detection map of the survey is shown in Fig. \ref{fig:det_lim_survey}.

As expected, the detectability function increases for larger masses and shorter periods. For periods between 1 d and 10 d, the average $p =  90 \%$ detection level corresponds to $m_p \sin i = 9.3$ M$_\oplus$, while for the longest considered periods, $[1000,3000]$ d, it corresponds to masses as large as  $m_p \sin i \simeq 180$ M$_\oplus$.

It is worth noticing that most of the planets and candidates are found in the region of the parameter space with intermediate detectability, around the $p = 50 \%$ level, while a few are found in regions with very low $p$ (around the targets with the largest and best-sampled time series). 

\section{Occurrence rates}
\label{sec:occurrence_rates}

Given the Detectability Function, $p$, the planetary occurrence rate, $f_\text{occ}$, can be computed as the average number of planets per star in a given region of the parameter space $\Delta(P,M)$ from the Poisson distribution:

\begin{equation}
\label{eq:poisson}
 \mathcal{P}(k \mid n, f_\text{occ}) = {(n f_\text{occ})^k e^{-n f_\text{occ}} \over {k!}},
\end{equation}

where $k = k_{\Delta(P,M)}$ is the number of planets detected in the chosen parameter space interval $\Delta(P,M)$, and the expected value is computed as the product between $n = n_{\Delta(P,M)}$, the number of targets sensitive to planets in $\Delta(P,M)$ , and $f_\text{occ}$. The number of sensitive targets, $n_{\Delta(P,M)}$, can be computed as the mean value of $p$ over $\Delta(P,M)$ multiplied by the number of stars in the survey $N$.\footnote{The derivation of $n_{\Delta(P,M)}$ assumes a log-uniform distribution of $p$ over the selected parameter-space intervals     $\Delta(P,M)$.}
We can compute the planetary occurrence rate from Eq. \ref{eq:poisson} by considering it as- the (un-normalised) posterior distribution of $f_\text{occ}$, for given values of $k$ and $N$.  Therefore best-fit values of planetary occurrence  can be computed as the medians and $68\%$ confidence intervals. In the case of $k = 0$, i.e. no detected planet in a given interval, the upper limit of the planetary occurrence rate was computed as the $68$th percentile of the distribution.

\begin{figure}
\centering
\includegraphics[width=\linewidth]{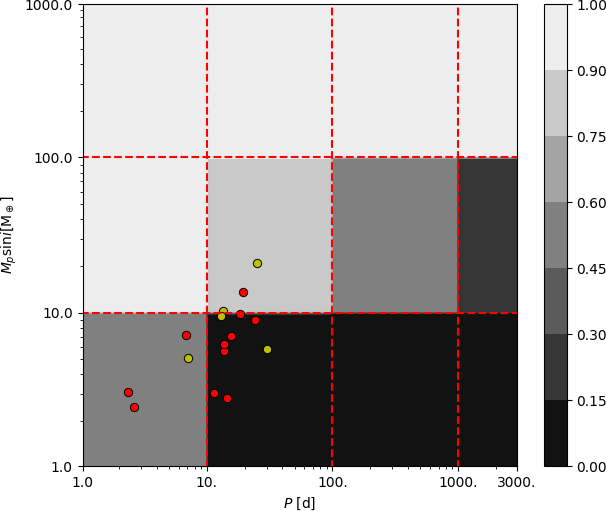}
\caption{Averaged HADES detection map. The color scale expresses the global detection function, $p$, averaged over the intervals $\Delta(P,M)$ defined for the computation of the occurrence rates. The red circles mark the position in the parameter space of the  confirmed HADES planets (see Fig. \ref{fig:planet_systems}), while the yellow circles indicate the additional candidates presented in this study (see Table \ref{tab:candidates}).}
\label{fig:det_lim_survey_binned}
\end{figure}

To compute the occurrence rates $f_\text{occ}$, we defined the following parameter space interval $\Delta(P,M)$: the $m_p \sin i$ was divided in three intervals, $[1.,10.]$ M$_\oplus$, $[10.,100.]$ M$_\oplus$, and $[100.,1000]$ M$_\oplus$; the $P$ axis was divided in four intervals, $[1,10]$ d, $[10,10^2]$ d, $[10^2,10^3]$ d, and $[10^3,3\times10^3]$ d.\footnote{We chose these large intervals, even if most of our detections are grouped at short periods, because these are the most common period and mass intervals used in the literature for computing the planetary occurrence rates. In particular, the two previous studies on M-dwarf planetary populations from RVs of \citet{bonfils13} and \citet{tuomi14} use these same definitions, thus easing the comparison of our results.} The averaged detection function $\langle p \rangle_{\Delta(P,M)}$ is shown in Fig. \ref{fig:det_lim_survey_binned}.

We computed the occurrence rates first using only the confirmed HADES planets, shown in Fig. \ref{fig:planet_systems}, inserting in Eq. \ref{eq:poisson} the number of confirmed planet per parameter space interval $k_p = k_{p,\Delta(P,M)}$. Then we considered also the candidate planets presented in Sect. \ref{sec:candidates}, thus computing the total number of both confirmed planets and candidates per parameter space interval $k_c = k_{c,\Delta(P,M)}$. The resulting occurrence rates derived from the confirmed-only and confirmed+candidates planets are listed in Table \ref{tab:focc_planets} and \ref{tab:focc_candidates} respectively.

\begin{table*}
\caption[]{Occurrence rates of planets in the HADES sample.}
\label{tab:focc_planets}
\centering
\begin{tabular}{l|cccc}
\hline
\hline
 & \multicolumn{4}{c}{Period} \\
$m_p \sin i$ & \multicolumn{4}{c}{$[$d$]$} \\
$[$M$_\oplus]$ & $[1,10]$ & $[10,10^2]$ & $[10^2,10^3]$ & $[10^3,3 \cdot 10^3]$ \\
\hline
$[10^2,10^3]$ & $k_p = 0$ & $k_p = 0$ & $k_p = 0$ & $k_p = 0$ \\
 - & $n = 56.0$ & $n = 55.8$ & $n = 54.4$ & $n = 51.7$ \\
 - & $f_\text{occ} < 0.02$ & $f_\text{occ} < 0.02 $ & $f_\text{occ} < 0.02$ & $f_\text{occ}  < 0.02$ \\
$[10,10^2]$ & $k_p = 0$ & $k_p = 1$ & $k_p = 0$ & $k_p = 0$ \\
 - & $n = 55.0$ & $n = 49.3$ & $n = 33.2$ & $n = 14.9$ \\
 - & $f_\text{occ} < 0.02$ & $f_\text{occ} = 0.02^{+0.05}_{-0.01}$ & $f_\text{occ} < 0.04$ & $f_\text{occ}  < 0.08$ \\
$[1,10]$ & $k_p = 3$ & $k_p = 7$ & $k_p = 0$ & $k_p = 0$ \\
 - & $n = 29.0$ & $n = 8.2$ & $n = 0.7$ & $n = 0.4$ \\
 - & $f_\text{occ} = 0.10^{+0.10}_{-0.03}$ & $f_\text{occ} = 0.85^{+0.46}_{-0.21}$ & $f_\text{occ} < 1.65 $ & $f_\text{occ} < 2.70$ \\
\hline
\end{tabular}
\tablefoot{$k_p$ and  $n$ are the number of detected planets and of stars sensitive to planets in the parameter space interval, respectively, and $f_\text{occ}$ is the planetary occurrence rate, computed as average number of planets per star.}
\end{table*}

\begin{table*}
\caption[]{Same as Table \ref{tab:focc_planets}, taking into account also the candidate planets announced in Sect. \ref{sec:candidates}.}
\label{tab:focc_candidates}
\centering
\begin{tabular}{l|cccc}
\hline
\hline
 & \multicolumn{4}{c}{Period} \\
$m_p \sin i$ & \multicolumn{4}{c}{$[$d$]$} \\
$[$M$_\oplus]$ & $[1,10]$ & $[10,10^2]$ & $[10^2,10^3]$ & $[10^3,3 \cdot 10^3]$ \\
\hline
$[10^2,10^3]$ & $k_p = 0$ & $k_p = 0$ & $k_p = 0$ & $k_p = 0$ \\
 - & $n = 56.0$ & $n = 55.8$ & $n = 54.4$ & $n = 51.7$ \\
 -  & $f_\text{occ} < 0.02$ & $f_\text{occ} < 0.02 $ & $f_\text{occ} < 0.02$ & $f_\text{occ}  < 0.02$ \\
$[10,10^2]$ & $k_p = 0$ & $k_p = 3$ & $k_p = 0$ & $k_p = 0$ \\
 - & $n = 55.0$ & $n = 49.3$ & $n = 33.2$ & $n = 14.9$ \\
 - & $f_\text{occ} < 0.02$ & $f_\text{occ} = 0.06^{+0.06}_{-0.02}$ & $f_\text{occ} < 0.04$ & $f_\text{occ}  < 0.08$ \\
$[1,10]$ & $k_p = 4$ & $k_p = 9$ & $k_p = 0$ & $k_p = 0$ \\
 - & $n = 29.0$ & $n = 8.2$ & $n = 0.7$ & $n = 0.4$ \\
 - & $f_\text{occ} = 0.14^{+0.11}_{-0.04}$ & $f_\text{occ} = 0.97^{+0.48}_{-0.23}$ & $f_\text{occ} < 1.65$ & $f_\text{occ} < 2.70$ \\
\hline
\end{tabular}
\end{table*}

\begin{figure}
   \centering
   \subfloat[][]
 {\includegraphics[width=.45\textwidth]{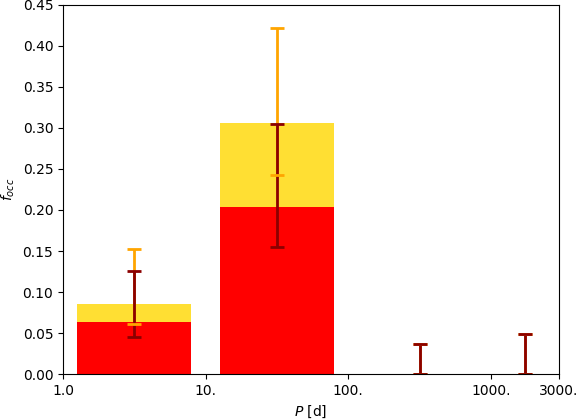}
 \label{fig:focc_dist_period:1d}} \\
   \subfloat[][]
 {\includegraphics[width=.45\textwidth]{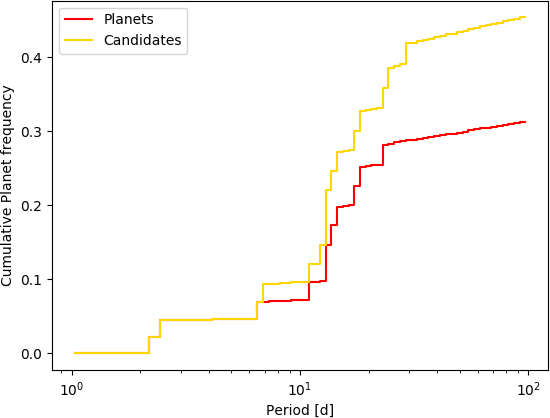}
 \label{fig:focc_dist_period:cum}}
 \caption{Planetary occurrence rate $f_\text{occ}$ as a function of the orbital period, with the corresponding 1$\sigma$ uncertainties (top panel). The median cumulative planet frequency is shown in the lower panel. The red and yellow distributions correspond to the occurrence rates derived from the confirmed-only, $k_p$, and confirmed+candidates planets, $k_c$, respectively.}
 \label{fig:focc_dist_period}
\end{figure}

\begin{figure}
   \centering
   \subfloat[][]
 {\includegraphics[width=.45\textwidth]{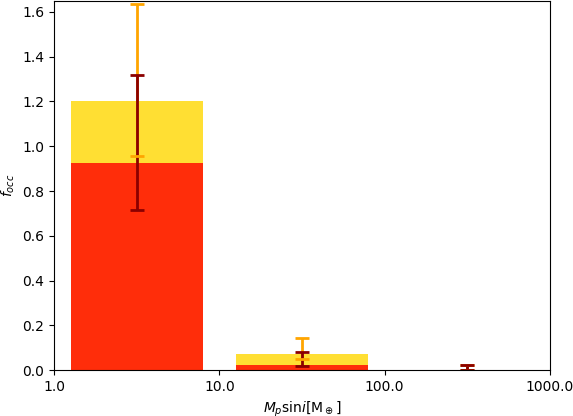}
 \label{fig:focc_dist_mass:1d}} \\
   \subfloat[][]
 {\includegraphics[width=.45\textwidth]{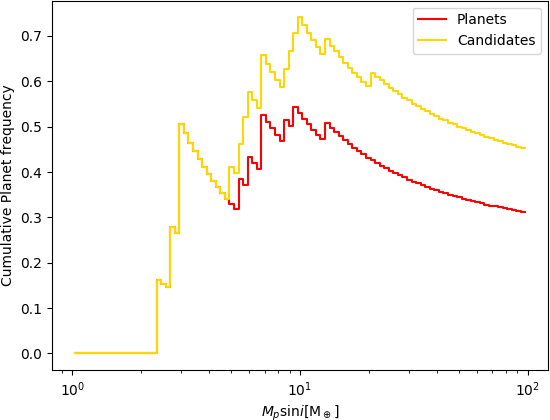}
 \label{fig:focc_dist_mass:cum}}
 \caption{Planetary occurrence rate $f_\text{occ}$ as a function of the minimum mass , with the corresponding 1$\sigma$ uncertainties (top panel). The median cumulative planet frequency is shown in the lower panel. The red and yellow distributions correspond to the occurrence rates derived from the confirmed-only, $k_p$, and confirmed+candidates planets, $k_c$, respectively.}
 \label{fig:focc_dist_mass}
\end{figure}

In addition to the 2D occurrence rates shown in Table \ref{tab:focc_planets} and \ref{tab:focc_candidates}, we computed the 1D occurrence rate distribution for both period and minimum-mass. This was done by using the same bins defined previously for the 2D map, but integrating over the whole range of the other parameter, thus computing the frequency of planets of all masses as a function of period, and of planets of all periods as a function of minimum mass. The results are shown in Fig. \ref{fig:focc_dist_period:1d} and \ref{fig:focc_dist_mass:1d}. Furthermore, we computed the cumulative planet frequency for both the period and minimum mass. This was done to show the finer structure of the planetary occurrence rates. In both cases, we restricted the computation over the intervals $[1,100]$ d and $[1,100]$ M$_\oplus$, to focus on the region of the parameter space in which there were detected planets and candidates. We then divided these interval in a fine grid of bins, in order to reduce as much as possible the number of planets in each bin. We used 80 bins for the period distribution, and 100 bins for the minimum mass distribution. The cumulative planetary frequency was computed as the occurrence rate, integrating the detection function from the minimum value to the value of each bin, marginalizing over the other parameter. The results are shown in Fig. \ref{fig:focc_dist_period:cum} and \ref{fig:focc_dist_mass:cum}.

\subsection{HZ occurrence rates}
\label{sec:hz_occurrence}

An aspect of great interest in the studies of occurrence rates of low-mass planets is the parameter $\eta_\oplus$, the frequency of low-mass habitable planets, i.e. planets with masses  $m_p \sin i < 10$ M$_\oplus$ orbiting at the right distance from their host star to allow the presence of liquid water on their surface, within the so-called Habitable Zone \citep[HZ,][]{kastingetal1993}. We compute the HZ limits following the recipe from \citet{kopparapuetal2013}, defining a conservative and an optimistic HZ: the conservative HZ is computed adopting the Runaway Greenhouse and Maximum Greenhouse coefficients for its inner and outer limits respectively; the optimistic limits of the  HZ, $a_\text{HZ,in}$ and $a_\text{HZ,out}$, are instead computed with the Recent Venus and Early Mars coefficients, respectively \citep{kopparapuetal2013erratum}.

Thus far no  planet has been confirmed inside the conservative or optimistic HZs of any HADES target (see Fig. \ref{fig:planet_systems}). However, we can still compute an upper limit of the frequency of habitable planets.
Thus we generated an additional detection map of the HADES samples, as a function of the minimum mass and of the position inside the HZ: this was done, for each target, by computing the detectability functions $p_i$ as in Sect. \ref{sec:detection_limits}, but the posterior of the period $P_\text{test}$ was first converted into the posterior of semi-major axis $a_\text{test}$ and then the semi-major axis was converted in HZ logarithmic scale, $a_\text{test,HZ}$:
\begin{equation}
\label{eq:hz_scale}
 \log{a_\text{test,HZ}} = (\log{a_\text{test}} - \log{a_\text{HZ,in}})/(\log{a_\text{HZ,out}}-\log{a_\text{HZ,in}}),
\end{equation}
this way $a_\text{test,HZ} =0$ corresponds to the inner edge of the optimistic HZ, $a_\text{HZ,in}$, and  $a_\text{test,HZ} =1$ to the outer edge, $a_\text{HZ,out}$. This conversion allow us to compare the HZs of different stars, which are located at different intervals of the semi-major axis parameter space.

\begin{figure}
\centering
\includegraphics[width=.8\linewidth]{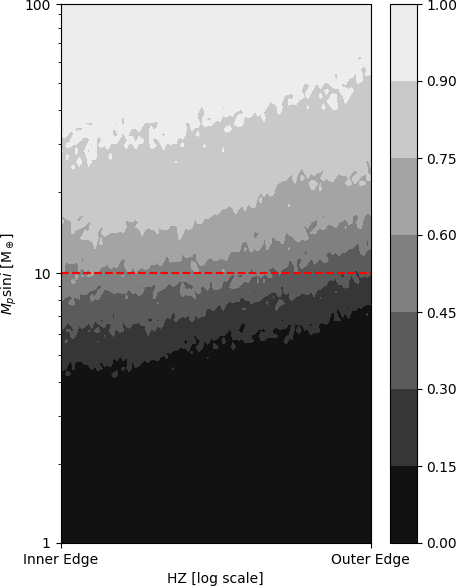}
\caption{Habitable Zone HADES detection map, derived as described in Sect. \ref{sec:hz_occurrence}. The color scale expresses the detection function, as in Fig. \ref{fig:det_lim_survey}.}
\label{fig:det_lim_HZ}
\end{figure}

For each star, we computed $\hat{p}_{\text{HZ},i}$ over a $100 \times 100$ logarithmic grid in the $[a_\text{HZ},M]$ parameter space, where the range of the converted semi-major axis, $a_\text{HZ}$, was $[0,1]$, and the minimum-mass range was $[1,100]$ M$_\oplus$. We then computed the global detection function inside the HZ $p_\text{HZ}$ following Eq. \ref{eq:glob_det_fun}.
The result is shown in Fig. \ref{fig:det_lim_HZ}.

From the detection function $p_\text{HZ}$, taking into account only the confirmed planets none of which were detected inside the HZ (i.e. $k_\text{HZ} = 0$), we could use Eq. \ref{eq:poisson} to compute the upper limits of the occurrence rate $f_\text{occ,HZ} = \eta_\oplus$ as before: for low-mass planets, $m_p \sin i < 10$ M$_\oplus$, we obtained an upper limit $f_\text{occ,HZ} = \eta_\oplus < 0.23$. Moreover, we computed also the occurrence rate of higher-mass planets inside the HZ, that is $10$ M$_\oplus < m_p \sin i < 100$ M$_\oplus$, obtaining $f_\text{occ,HZ} < 0.03$.

Moreover, we detected one potential planetary signal located inside the optimistic HZ of its host star: GJ 399 32.9 d candidate is located inside the HZ close to its inner-edge. However, this signal is slightly less significant than the threshold we adopted in Sect. \ref{sec:model_selection} ($\Delta \text{BIC} = 9.5$), and thus is not considered as a planetary candidate in Sect. \ref{sec:candidates} and Table \ref{tab:candidates}. Nonetheless, we chose to take into account to compute a second estimate of $\eta_\oplus$. Taking this candidate into account, the occurrence rate for $m_p \sin i < 10$ M$_\oplus$ increases to $f_\text{occ,HZ} = 0.20^{+0.45}_{-0.06}$.

\subsection{Long-period companions occurrence rates}
\label{sec:occurrence_trends}

It is more difficult to estimate the occurrence rates of long-period planets, i.e. $P \gtrsim 3000 $d, which could have been detected only as linear RV trends due to the relatively short timespan of our survey. Since in fact, as discussed in Sect. \ref{sec:trends}, we did detect some linear trends with no apparent stellar origin we discuss a few approximations and caveats adopted in order to try and estimate the associated long-period planets occurrence rates. It should be noticed that the uncertainties of such an estimate are intrinsically very large, and very strong approximations are required. Nonetheless the results could bring some interesting information on the little known field of long-period planets around M dwarfs, and thus we shall proceed anyway.

First we need to estimate the planetary parameters from the information we can obtain from the linear trends: an estimate of the minimum orbital period $P_\text{min}$ could be computed, in the approximation of circular orbit, as four times the timespan of the observations $P_\text{min} \gtrsim 4 \cdot T_\text{s}$\footnote{This can be assumed since the linear trend could cover up to half of the peak-to-trough part of a sinewave, without showing any evident curvature. Thus the observed timespan could correspond to no more than one fourth of the orbital period.}; similarly, the minimum RV semi-amplitude $K_\text{min}$ could be computed as the total RV variation during the observations, that is $K_\text{min} \gtrsim T_\text{s}\cdot  d$. From $P_\text{min}$ and $K_\text{min}$ the corresponding `minimum' minimum mass $(m_p \sin i)_\text{min}$ can be derived. The resulting minimum planetary parameters are listed in Table \ref{tab:minimum}. In the case of the GJ 3997 quadratic trend, the minimum RV amplitude $K_\text{min}$ could be approximated as the difference between the maximum and minimum RV values of the quadratic curve during the timespan of the observations, $K_\text{min} = RV_\text{max} - RV_\text{min}$, while the period could be computed as a function of the quadratic acceleration, $q$, as $P_\text{min} \gtrsim 2 \pi \sqrt{K_\text{min} / q}$ \citep{kippingetal2011}.

\begin{table}
\caption[]{Long-period candidate planetary signals assumed from the modelled linear RV trends.}
\label{tab:minimum}
\centering
\begin{tabular}{lcccc}
\hline
Target & $P_\text{min}$ & $a_\text{min}$ & $(m_p \sin i)_\text{min}$ & $n_t$ \\
 & $[$d$]$ &  $[$AU$]$ & $[$M$_\oplus]$ & \\
\hline
\noalign{\smallskip}
GJ 16 & $ 9000$ & $6.6$  & $140.$ & $26$ \\
\noalign{\smallskip}
GJ 15A & $ 7600$ & $5.5$ & $49.$ & $5$ \\
\noalign{\smallskip}
V$^*$BRPsc & $ 9000$ & $6.1$ & $95.$ & $14$ \\
\noalign{\smallskip}
NLTT 21156 & $ 8500$ & $6.5$ & $620.$ & $55$ \\
\noalign{\smallskip}
GJ 793 & $ 5000$ & $3.9$ & $59.$ & $34$ \\
\noalign{\smallskip}
GJ 412A & $ 6200$ & $4.8$ & $72.$ & $23$ \\
\noalign{\smallskip}
GJ 408 & $ 8200$ & $5.4$ & $74.$ & $14$ \\
\noalign{\smallskip}
GJ 4092 & $ 7700$ & $6.4$ & $170.$ & $35$ \\
\noalign{\smallskip}
\hline
\end{tabular}
\end{table}

Another obstacle to overcome is the computation of the detection efficiency for such long-period signals, which cannot be computed following our standard recipe described in Sect. \ref{sec:stat_analysis}, since the emcee test-planet approach cannot be robustly applied to periods much longer than the timespan of the time series. We decided to assess the detection efficiency of the HADES time series towards long-period signals, i.e. $P \gtrsim 3000$ d, by computing for each detected long-term trend $d$ the number of time series in which that trend could be detected with a significance larger than $>3\sigma$, that is in which  $d > 3 \, \sigma_{d,i}$, where $\sigma_{d,i}$ is the measured uncertainty of the acceleration term in the best-fit MCMC solution of the $i$th time series. The resulting numbers of sensitive time series for each trend, $n_t$, are listed in the last column of Table \ref{tab:minimum}. The mean value of $n_t$ can be then used along the number of significant trends, $k_t$, to estimated the occurrence rate $f_{\text{occ},t}$ from Eq. \ref{eq:poisson}.

As discussed in Sect. \ref{sec:trends}, we conducted a few tests to ascertain if the measured RV trends could be caused by other phenomena such as magnetic cycles or stellar companions. A more in-depth analysis is beyond the scope of the present work, but we can compute some preliminary estimates of the occurrence rates, assuming that all the observed trends are due to actual long-period planetary companions. The resulting frequency of planets with $P \gtrsim 3000$ d and $m_p \sin i \gtrsim 10$ M$_\oplus$ is thus $f_{\text{occ},t} = 0.31^{+0.15}_{-0.07}$. Since the minimum masses listed in Table \ref{tab:minimum} are quite diverse, and cover two different intervals of masses defined in the analysis of the short period planet occurrence rates (Fig. \ref{fig:det_lim_survey_binned}), we could divide the trend sample in two subsamples with minimum masses larger and smaller than $100$ M$_\oplus$. Computing separately the occurrence rates for these two subsamples, we obtain $f_{\text{occ},t} = 0.08^{+0.08}_{-0.02}$ for long-period planets with $m_p \sin i > 100$ M$_\oplus$, and $f_{\text{occ},t} = 0.28^{+0.19}_{-0.08}$ for long-period planets with $10$ M$_\oplus < m_p \sin i < 100$ M$_\oplus$.

\section{Discussion}
\label{sec:discussion}
We will now discuss the main features of our results concerning the sensitivity of the HADES survey and the occurrence rates of extrasolar planets around early M dwarfs, comparing them with previous studies on the planetary population of M dwarfs and earlier-type stars, both from RV surveys and photometric surveys such as Kepler. It is worth noticing that the comparison between RV and photometric surveys is far from obvious, since it strongly depends on the mass-radius relations which are still largely unknown, in particular for small low-mass planets.

\subsection{Comparison with other RV surveys}

We can see in Fig. \ref{fig:focc_dist_mass:1d} that the occurrence rate of planets around early-M dwarfs increases strongly towards lower masses. If we take into account only the confirmed HADES planets (red distribution), the occurrence rate rises from $f_\text{occ} = 0.02^{+0.06}_{-0.01}$, for $10$ M$_\oplus < m_p \sin i < 100$ M$_\oplus$, to $f_\text{occ} = 0.93^{+0.39}_{-0.21}$, for $1$ M$_\oplus < m_p \sin i < 10$ M$_\oplus$.  This confirms the expected behaviour that has been previously observed in other surveys of later-M dwarfs \citep{bonfils13,tuomi14}. The occurrence rate as a function of the orbital period, as shown in Fig. \ref{fig:focc_dist_period:1d}, instead appears to peak at intermediate periods, with the highest value $f_\text{occ} = 0.20^{+0.10}_{-0.05}$, for $10$ d$ < P < 100$ d, which is roughly $2\sigma$ higher than the value for $P < 10$ d, $f_\text{occ} = 0.06^{+0.06}_{-0.02}$. These behaviours are accentuated if we take into account also the additional planetary candidates presented in Sect. \ref{sec:candidates} (yellow distributions in Fig. \ref{fig:focc_dist_period}). It is worth noticing that this is due to the fact that most of the detected planets and candidates in the HADES samples are gathered around periods of a few tens of days and masses of around $10$ M$_\oplus$ (Fig. \ref{fig:det_lim_survey}), even if the detection probability in that region of the parameter space is quite low, $p \simeq 50 \%$. This confirms what was previously reported by \citet{tuomi14} in their study of UVES and HARPS RV data, and could also correspond to the overabundance of transiting planets with similar periods and radii smaller than $3$ R$_\oplus$ observed around Kepler M dwarfs: \citet{dressingcharbonneau2015} observed a higher occurrence rate of planets with $10$ d$ < P < 50$ d with respect to those with $P < 10$ d, even if not as significant as observed in our sample. Moreover, it is worth noticing that the cumulative planet distribution as a function of the minimum mass (Fig. \ref{fig:focc_dist_mass:cum}) shows hints that the planet mass distribution might show a valley between 3 and 5 M$_\oplus$, where the cumulative frequency decreases steadily due to the increasing sensitivity of the survey. However, due to the limited number of detected planets, the uncertainties on the fine details of the planet distributions are still quite large.

As shown in Table \ref{tab:focc_planets}, we derive an occurrence rate of low-mass ($1$ M$_\oplus < m_p \sin i < 10$ M$_\oplus$) short-period ($1$ d$ < P < 10$ d) planets around early-M dwarfs of $f_\text{occ} = 0.10^{+0.10}_{-0.03}$, which is lower ($\simeq2\sigma$) than the frequency derived by \citet{bonfils13} for later-M dwarfs, $f_\text{occ} = 0.36^{+0.24}_{-0.10}$. This could indicate that early and late type M dwarfs have different population of low-mass short-period planets. Even taking into account all the yet-to-be-confirmed candidates planets we discussed in Sect. \ref{sec:candidates}, our estimates of $f_\text{occ}$ is still $\simeq 1.5\sigma$ lower than that of \citet{bonfils13} (Table \ref{tab:focc_candidates}), which could still be evidence of this phenomenon. On the other hand, the same effect does not appear to be present for periods $10$ d$ < P < 100$ d, where our estimate of $f_\text{occ} = 0.85^{+0.46}_{-0.21}$ is perfectly compatible with the value obtained for later-M dwarfs, or even higher if all the candidates planets were confirmed (although still within the $1\sigma$ uncertainty of the value obtained by \citet{bonfils13} $f_\text{occ} = 0.52^{+0.50}_{-0.16}$). It is worth noticing that we find very large upper limits ($>1$) for low-mass planets with periods larger than $100$ d, and this is in line with previous results \citep{tuomi14}: this is due to the low sensitivity of RV surveys to low-mass planets at these periods. It is however interesting to notice that, due to this lack of observational constraints, low-mass planets at intermediate periods ($100$ d$ < P < 1000$ d) could be very abundant, even if to date only four such planets have been detected between both RV and transit surveys\footnote{NASA exoplanet archive - 08/04/2021: GJ 180\,d $P = 106.300 \pm 0.129$ d, $m_p \sin i = 7.56 \pm 1.07$ M$_\oplus$, GJ 229A\,c $P = 121.995 \pm 0.161$ d, $m_p \sin i = 7.268 \pm 1.256$ M$_\oplus$ \citep{fengetal2020}; GJ 628\,d $P = 217.21^{+0.55}_{-0.52}$ d, $m_p \sin i = 7.70^{+1.12}_{-1.06}$ M$_\oplus$ \citep{astudillodefruetal2017}; GJ 667C\,g $P = 256.2^{+13.8}_{-7.9}$ d, $m_p \sin i = 4.6^{+2.6}_{-2.3}$ M$_\oplus$ \citep{angladaescudeetal2013}.}.

\citet{sabottaetal2021}, analysing a sub-sample of the CARMENES survey, found occurrence rates slightly higher than \citet{bonfils13}, even though compatible within 1$\sigma$ with their results in all period and minimum-mass bins in which both survey had detected planets. It is interesting to notice that \citet{sabottaetal2021} found a relatively high frequency of giant planets ($100$ M$_\oplus < m_p \sin i < 1000$ M$_\oplus$) with periods up to 1000 d, $f_\text{occ} = 0.06^{+0.04}_{-0.03}$, which is somewhat larger than our upper limit of $f_\text{occ} < 0.02$. This however could be explained since, as they mention in their work, their results could be biased by the selection criteria of the analysed sub-sample.
For intermediate masses ($10$ M$_\oplus < m_p \sin i < 100$ M$_\oplus$), \citet{sabottaetal2021} found increasing planetary frequencies towards longer periods up to 1000 d, which appears to be in contrast with our occurrence rates that peaks at intermediate periods (see Table \ref{tab:focc_planets}); it is however worth noticing that our estimate of frequency of long-period planets (see Sect. \ref{sec:disc_longper}) is quite large, which could confirm the increasing trend observed by \citet{sabottaetal2021}. Looking at the frequency of low-mass short-period planets, \citet{sabottaetal2021} found $f_\text{occ} = 0.59^{+0.20}_{-0.17}$, higher than both our estimates and the one from the HARPS M-dwarf survey \citep{bonfils13}. This again could be explained by the difference in spectral type of the three samples, as the sub-sample analysed by \citet{sabottaetal2021} includes later-M dwarfs that are not observed by the other surveys. Moreover, considering separately stars with masses higher and lower than $0.34$ M$_\odot$, the CARMENES occurrence rates become $f_\text{occ} < 0.22$ and $f_\text{occ} = 1.06^{+0.35}_{-0.28}$ respectively: while the former value is compatible with our estimate of $f_\text{occ} = 0.10^{+0.10}_{-0.03}$, the latter is significantly higher ($>3\sigma$) than our value, which again suggests that the frequency of small-mass planets strongly increases with later spectral types.

\subsection{Comparison with transiting planets}
 
Comparing with the frequency of small planets around Kepler M dwarfs, it is evident that our derived occurrence rates for early-M dwarfs are much lower than the average per star of $2.2\pm 0.3$ planets with $1$ R$_\oplus < R_p < 4$ R$_\oplus$ and $1.5$ d$ < P < 180$ d derived by \citet{gaidosetal2016}.\footnote{Other studies of Kepler M dwarfs, found lower number of planets per star, $\sim 1.5$ for $1$ R$_\oplus < R_p < 4$ R$_\oplus$ and $1.$ d$ < P < 10$ \citep[and references therein]{sabottaetal2021}, however these numbers are still much higher than the occurrence rates we derive from the HADES RV sample.} However, \citet{hardegreeullmanetal2019} found that the number of planets with orbital periods shorter than 10 d per stars for Kepler mid-M dwarfs decreases from M5 V to M3 V stars. Assuming this effect applies also to intermediate-period planets, this could mitigate the difference between our RV-based results and Kepler occurrence rates, as the sample analysed by \citet{gaidosetal2016} includes many later-type M dwarfs which are not included in our survey. The results from \citet{hardegreeullmanetal2019} also confirm the discrepancy we observed between our early-M sample and the later-type survey of \citet{bonfils13}.  

\begin{figure}
\centering
\includegraphics[width=.8\linewidth]{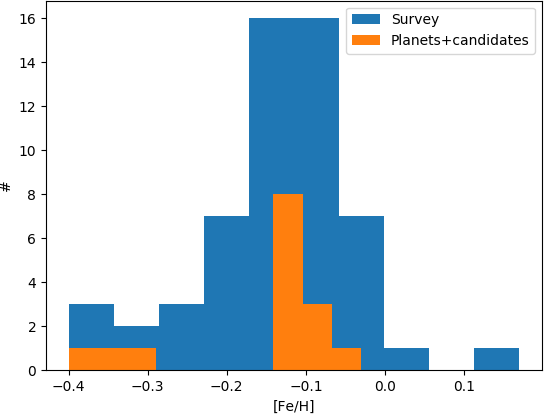}
\caption{Distribution of stellar metallicities for the HADES samples and for the sub-sample of planet-hosting stars (including also planetary candidates from Sect. \ref{sec:candidates}).}
\label{fig:metallicity_planets}
\end{figure}

Moreover, \citet{muirheadetal2015} found that $0.16^{+0.02}_{-0.02}$ of Kepler early-M dwarfs host compact multi-planets systems with orbital periods lower than 10 d: while this frequency could be consistent with the frequency of short-period low-mass planetary systems we derive from our survey, $f_{\text{occ}} = 13.8^{+9.8}_{-4.1}\%$ taking into account also the candidate planets\footnote{Assuming the empyrical planet mass–radius relationships of \citet{marcyetal2014} adopted by \citet{muirheadetal2015}.}, we found only one system hosting two short period planets \citep[GJ 3998][]{afferetal2016}, while all other detected systems contained only a single planet.
This apparent incompatibility could be mitigated by selection effect, since many multiple systems discovered by Kepler are composed of small 1-2 R$_\oplus$ planets, which could easily fall below our $2.8^{+3.9}_{-1.3}$ M$_\oplus$ detection threshold for planets with period shorter than 10 d \citep{marcyetal2014}. Moreover, many Kepler multi-planet systems host planets with very short periods below $2$ d, which are inherently difficult to detect in ground-based surveys due to daily sampling limitations.
Another possible explanation of the discrepancy between the number of multi-planet systems found in our RV survey and the statistics of Kepler planets, could be found in the metallicity of our sample: recent studies \citep[e.g.][]{andersonetal2021} found evidence that Kepler compact multiple planetary systems are hosted by M dwarfs statistically more metal-poor than the general population of systems with no detected transiting planets. This is coherent with the observed behaviour of Sun-like stars, which show increasing frequencies of compact multiple systems for metallicities $[$Fe$/$H$] < -0.3$ \citep{breweretal2018}. The complete HADES sample has a mean $[$Fe$/$H$] = -0.13$, and the planet-hosting subsample has a similar mean metallicity of $[$Fe$/$H$] = -0.15$ (see Fig. \ref{fig:metallicity_planets}): since the M dwarfs in our sample are relatively metal-rich compared to the bulk of compact-system hosting Kepler stars, this could explain the scarcity of multiple planetary systems detected by our survey.
This could also be in line with the results from the analysis in \citet{maldonadoetal2020}, that showed that there might be an anti-correlation between the frequency of low-mass planets and stellar metallicity.

\subsection{Long-period and giant planets}
\label{sec:disc_longper}

From our analysis of long-term trends, in the assumption that all the trends listed in Table \ref{tab:trends} are in fact of planetary origin, we can estimate the frequency of long-period ($P > 3 \cdot 10^3$ d) planets. We estimate that, for intermediate masses ($10$ M$_\oplus < m_p \sin i < 100$ M$_\oplus$), such planets could be quite abundant around early-M dwarfs, $f_{\text{occ},t} = 0.28^{+0.19}_{-0.08}$, with a much higher frequency than the estimate for the same masses and periods $10^3$ d$ < P < 3 \cdot 10^3$ d (Table \ref{tab:focc_candidates}). This result could be confirmed by the similar occurrence rate $f_\text{occ} \simeq 0.20$ computed by \citet{tuomi14} for planets with periods longer than 1000 d, even if their survey suffered from the same limitations discussed in Sect. \ref{sec:occurrence_trends} due to the similar timespan of the observations. Moreover, this can confirm the trend observed by \citet{sabottaetal2021}, who measured increasing intermediate-mass planet frequencies at incresing periods.
Thus our survey confirms the scarcity of giant planets ($100$ M$_\oplus < m_p \sin i < 1000$ M$_\oplus$) at short-to-intermediate periods around M dwarfs, setting an upper limit to their occurrence $f_\text{occ} < 0.02$ for periods $P < 3000$ d (Table \ref{tab:focc_planets}). This is consistent with the behaviour observed for later-M dwarfs by \citet{bonfils13}, who estimated the frequency of giant planets of periods up to $1000$ d to be $f_\text{occ} \simeq 0.02$, even if their sample did in fact contain 2 of such planets. Moreover, having detected 2 very-long period giant planets (GJ 849\,b and GJ 832\,b), they estimated the frequency of giant planets with periods $10^3$ d$ < P < 10^4$ d to be  $f_\text{occ}  = 0.04^{+0.05}_{-0.01}$. Even if the timespan of our survey did not allow for the detection of such long-period planets, this result is perfectly compatible with our estimate of the frequency of long-period giant planets derived from the observed RV trends $f_{\text{occ},t} = 0.08^{+0.08}_{-0.02}$ (Sect. \ref{sec:occurrence_trends}). However, the occurrence rate we derive is larger ($\simeq 2\sigma$) than the estimated frequency of Jupiter-like planets around M dwarfs computed combining RV and microlensing surveys, $f_J = 0.03^{+0.01}_{-0.02}$ \citep{clantongaudi2014}. Moreover, it is worth noticing that the $f_{\text{occ},t}$ we derived is very close with the recent results by \citet{wittenmyer2020}, who derived the frequency of cool Jupiters to be $0.07^{+0.02}_{-0.01}$ from an RV survey of FGK stars. This could suggest that the frequency of long-period giant planets does not vary as strongly as the frequency of low-mass planets from Solar-type stars to M dwarfs.

Comparing Fig. \ref{fig:planet_systems} and Table \ref{tab:candidates} to Table \ref{tab:trends}, it is interesting to notice that, apart from GJ 15A, none of our stars hosting short-period low-mass planets shows also a long-period trend compatible with the presence of an outer planetary companion. This appears to be in contrast with theoretical models, since \citet{izidoroetal2015} predicted that systems with only one low-mass planet with orbital period shorter than 100 d should also harbour a Jupiter-like planet on a more distant orbit: otherwise both in situ formation and inward migration models predict that super-Earth and warm Neptunes should form in rich systems hosting many close-in planets. However, of all of our detected planets and candidates, only one system appears to host more than a single close-in planet (GJ 3998), and, as previously mentioned, the other systems do not show any evidence of the expected outer massive companions.

\subsection{Habitable planets}

The HADES program has not been able to confirm the presence of any planet in the HZ so far, and thus, even if the detection function was generally low for low-mass planets in the HZ (Fig. \ref{fig:det_lim_HZ}), we derived an upper limit to the frequency of habitable planets around early-M dwarfs of $\eta_\oplus < 0.23$ (see Sect. \ref{sec:hz_occurrence}).
\citet{bonfils13} estimated a higher frequency of habitable planets around M dwarfs $\eta_\oplus = 0.41^{+0.54}_{-0.13}$\footnote{\citet{bonfils13} used the definition of HZ from \citet{selsisetal2007}, while \citet{tuomi14} used the same updated definition we adopted in our work \citep{kopparapuetal2013}.}. However, this value takes into account the presence of the habitable planet GJ 581\,d, which was later discarded as an artifact of stellar activity \citep{baluev2013b,robertsonetal2014}. Thus, recomputing their value of $\eta_\oplus$ considering only the other habitable planet detected in their survey, the obtained value is lower $\eta_\oplus \simeq 0.20$, which is much closer to the value derived from our analysis of the HADES survey. Moreover, \citet{tuomi14} found a similar value of $\eta_\oplus = 0.21^{+0.03}_{-0.05}$ considering planets with masses $3$ M$_\oplus < m_p \sin i < 10$ M$_\oplus$, which is compatible within $1\sigma$ with our estimate. The upper limit we derived, $\eta_\oplus < 0.23$, could still allow a higher number of low-mass planets in the HZ of early-M dwarfs compared to G-type stars, for which the fraction of Earth-like planets is estimated to be $\eta_\oplus \simeq 0.10$ \citep{perryman2018}. Instead, if we consider the weak candidate planet detected in this analysis orbiting in the outer rim of the HZ of its host star, our estimate of $\eta_\oplus$ increases to $\eta_\oplus = 0.20^{+0.45}_{-0.06}$, which would be compatible with the previous estimates from RV M-dwarf surveys, and confirm the abundance of habitable planets around low-mass stars compared to Solar-type stars.

\section{Conclusions}
\label{sec:conclusions}

We presented the statistical analysis of the spectroscopic observations of a magnitude-limited sample of nearby early-M dwarfs, observed within the HADES programme. The sample is composed of 56 targets, with an average of 77 high-precision RVs each. 
All time series in the sample have been analysed with a uniform Bayesian technique, in order to get consistent and unbiased estimates of the detection limits and planetary occurrence rates. Moreover, we applied GP regression to refine the planetary parameters and improve the detection efficiency around the most observed and active targets in the survey.

The sample includes 10 published planetary systems discovered and characterized as part of the HADES programme, and in this work we present 5 new planetary candidates.
Moreover we discuss 8 RV long-term trends that could be produced by long-period planetary companions.
The new candidates will be further analysed and discussed in future focused publications, two of which are in preparation and soon to be submitted (Gonz\'alez Hern\'andez et al. on GJ 21 and Affer et al. on GJ 3822). The other candidates, and in particular the weak HZ candidate discussed in Sect. \ref{sec:hz_occurrence}, will need additional RV observations to confirm or disprove their planetary nature. Similarly, an in-depth analysis of the long-term trends will be the focus of a future dedicated work, as a survey focused on the  characterization of late-type systems hosting long-period planets is currently being carried out within the GAPS programme \citep{barbatoetal2020}.

We confirm that giant planets ($100$ M$_\oplus < m_p \sin i < 1000$ M$_\oplus$) are very rare around M dwarfs at short-to-intermediate periods ($P < 3000$ d), $f_\text{occ} <0.02$. On the other hand, low-mass planets ($1$ M$_\oplus < m_p \sin i < 10$ M$_\oplus$) appear to be common, with frequencies varying between $f_\text{occ} = 0.10^{+0.10}_{-0.03}$ for short periods ($1$ d$ < P < 10$ d) to $f_\text{occ} = 0.85^{+0.46}_{-0.21}$ for longer periods ($10$ d$ < P < 100$ d). While high compared to Solar-type stars, these frequencies appear to be lower than for later-M systems, thus confirming the strong dependence of planetary occurrence rates on stellar mass. It is worth noticing that  early- and late-M dwarfs have very different internal structures, as the latter are fully convective, and this affects the efficiency of tidal interaction between a star and a close-by planet \citep{barker2020}, which could explain the significant difference in planetary occurrence rates between early- and late-M dwarfs. We also estimated the frequency of habitable planets around early-M dwarfs, finding an upper limit of $\eta_\oplus < 0.23$, which suggests that the frequency of Earth-like planets is as high if not higher than around Solar-type stars.

Finally, it is worth noticing how the current number of HADES close-in planets (11-16) is incompatible with the prediction of  $3.8 \pm 1.9$ detections \citep{pergeretal2017}, based on previous M-dwarf planet population models. This shows the importance of long-term high-precision surveys focused on narrow intervals of stellar masses, to improve our knowledge of planetary populations, and thus formation mechanisms, through different classes of stellar hosts.

\begin{acknowledgements}
M.Pi., A.P., A.M., and A.S. acknowledge partial contribution from the agreement ASI-INAF n.2018-16-HH.0. M.Pi. acknowledges  partial contribution from OB.FU. 1.05.02.85.13 ``Planetary Systems At Early Ages (PLATEA)''. J.M., and G.M. acknowledge support from the accordo attuativo ASI-INAF n.2021-5-HH.0 ``Partecipazione italiana alla fase B2/C della missione Ariel''.
E.G.A. acknowledge support from the Spanish Ministery for Science, Innovation, and Universities through projects AYA-2016-79425-C3-1/2/3-P, AYA2015-69350-C3-2-P, ESP2017-87676-C5-2- R, ESP2017-87143-R. The Centro de Astrobiología (CAB, CSIC-INTA) is a Center of Excellence ``Maria de Maeztu''. M.D. acknowledges financial support from the FP7-SPACE Project ETAEARTH (GA no. 313014).
P.G. gratefully acknowledges support from the Italian Space Agency (ASI) under contract 2018-24-HH.0.
M.Pe. and I.R. acknowledge support from the Spanish Ministry of Science and Innovation and the European Regional Development Fund through grant PGC2018-098153-B- C33, as well as the support of the Generalitat de Catalunya/CERCA programme. J.I.G.H. acknowledges financial support from Spanish Ministry of Science and Innovation (MICINN) under the 2013 Ramón y Cajal program RYC-2013- 14875. B.T.P. acknowledges Fundación La Caixa for the financial support received in the form of a Ph.D. contract. A.S.M. acknowledges financial support from the Spanish MICINN under the 2019 Juan de la Cierva Programme. B.T.P., A.S.M., J.I.G.H. and R.R. acknowledge financial support from the Spanish MICINN project PID2020-117493GB-I00, and from the Government of the Canary Islands project ProID2020010129.\\
The HARPS-N Project is a collaboration between the Astronomical Observatory of the Geneva University (lead), the CfA in Cambridge, the Universities of St. Andrews and Edinburgh, the Queen’s University of Belfast, and the TNG-INAF Observatory. This work has made use of data from the European Space Agency (ESA) mission Gaia (\url{https://www.cosmos.esa.int/gaia}), processed by the Gaia Data Processing and Analysis Consortium (DPAC, \url{https://www.cosmos.esa.int/web/gaia/dpac/consortium}). Funding for the DPAC has been provided by national institutions, in particular the institutions participating in the Gaia Multilateral Agreement.
We also thank the anonymous referee for the insightful review.

\end{acknowledgements}

  \bibliographystyle{aa} 
  \bibliography{biblio}

\begin{appendix}

\section{MCMC analysis: Planetary parameters, GP hyper-parameters, and adopted Priors}
\label{app:priors}

We adopted uninformative priors for all the parameters and hyper-parameters in our MCMC models. This was done for consistency across the statistical analysis, and to avoid adulterating the computed detection function by adopting incorrect informative priors. For most of the RV model parameters and GP hyper-parameters we adopted a uniform prior, except for the correlation decay timescale, $\lambda$, for which we adopted a uniform prior in logarithmic scale to avoid oversampling the long scales, since $\lambda$ can be of the order of tens or hundreds of days and more, depending on the activity level of the target. When an eccentric Keplerian signal was included in the model, instead of fitting separately $e_j$ and $\omega_j$, we use the auxiliary parameters $C_j = \sqrt{e_j}\cdot \cos \omega_j$ and $S_j = \sqrt{e_j} \cdot \sin \omega_j$ to reduce the covariance between $e_j$ and $\omega_j$, especially for low eccentricity value. Moreover, when the eccentricity was consistent with $e_j = 0$ we adopted a circular model, fixing $e_j = \omega_j = 0$, to reduce the number of parameters and the computational weight of the analysis.

The analysis of GJ 15A required some different priors for the RV model since, as discussed in \citet{pinamontietal2018}, the complete orbital solution was obtained by a combination of HIRES + HARPS-N data, and was not recoverable from HARPS-N data only. For this reason, we applied Gaussian priors to the GP hyper-parameters, to the semi-amplitude and period of the inner planets GJ 15A\,b, and to the acceleration coefficient $d$, which as discussed in Sect. \ref{sec:signals} corresponds to the RV variations caused by the outer planet GJ 15A\,c over the timespan of the HARPS-N observations.

Finally, particular care was required for the three parameters of the \textit{test-planet} RV model, whose prior could directly impact the resulting $\hat{p}_i (P,M)$ function. For the orbital period $P_\text{test}$ we adopted a uniform prior in logarithmic scale over the interval $\log \mathcal{U}$(1,3000) d, choosing the upper limit to be roughly twice the average timespan of the observations. The prior of RV amplitude $K_\text{test}$ required a particularly large prior, to avoid constraining the derived minimum-mass values over all the explored orbital period values: for this reason we adopted a very broad logarithmic uniform prior $\log \mathcal{U}$(0.01,100) m s$^{-1}$, with an upper limit sufficiently larger than the maximum r.m.s. of the HADES time series (see Table \ref{tab:star_pam}). Lastly, to fit the reference time of the \textit{test-planet} orbit, $T_{0,\text{test}}$, we defined the auxiliary parameter $t_{p,\text{test}} = T_{c}/P_\text{test}$, with $T_{c}$ the time of inferior conjunction: $t_{p,\text{test}}$ could be thus defined over a uniform prior $\mathcal{U}$(0,1),  avoiding the multi-modal distribution of $T_{c}$, which could greatly impact the efficiency of \texttt{emcee} \citep{foreman13}.

\begin{figure}
\centering
\includegraphics[width=\linewidth]{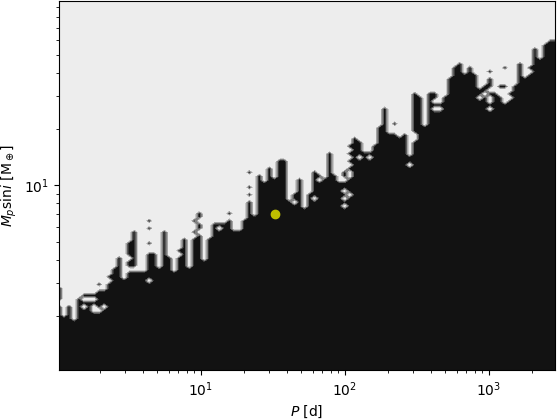}
\caption{Detection function map of the RV time series of GJ 399 (as in Fig. \ref{fig:det_lim_gj15a}). The yellow circle shows the position of the 32.9 d weak candidate signal.}
\label{fig:appendix_GJ399_detmap}
\end{figure}

Moreover, in Fig. \ref{fig:appendix_GJ399_detmap} is shown the detection function map of GJ399, with marked the position of the 32.9 signal identified in the RV time series which, as discussed in Sect. \ref{sec:hz_occurrence}, is just below the acceptance threshold $\Delta \text{BIC} = 10$: it is worth noticing that the weak-candidate signal mass, $M_p \sin i = 7.0 $ M$_\oplus$, is just below the $11.7^{+2.4}_{-4.7}$ M$_\oplus$ detection threshold for periods in $[20,40]$ d. This confirms the good correspondence between the adopted planetary acceptance criterion (Sect. \ref{sec:model_selection}) and the MCMC-derived detectability function (Sect. \ref{sec:algorithm}).

\section{New planetary candidates}
\label{app:candidates_plan}

\begin{figure*}
   \centering
   \subfloat[][]
 {\includegraphics[width=.45\textwidth]{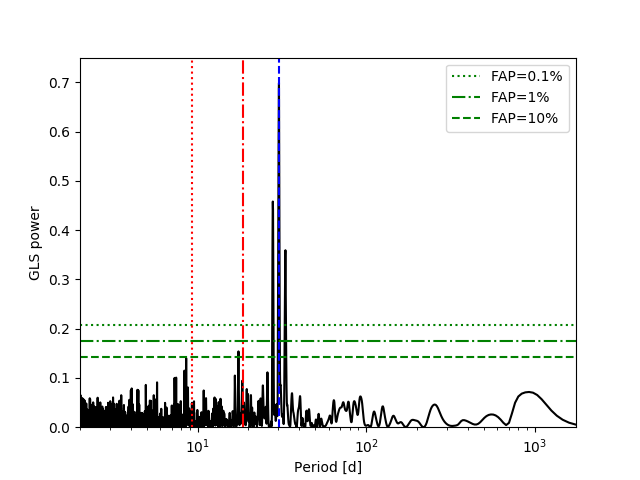}
 \label{fig:appendix_m006:periodogram}}
   \subfloat[][]
 {\includegraphics[width=.45\textwidth]{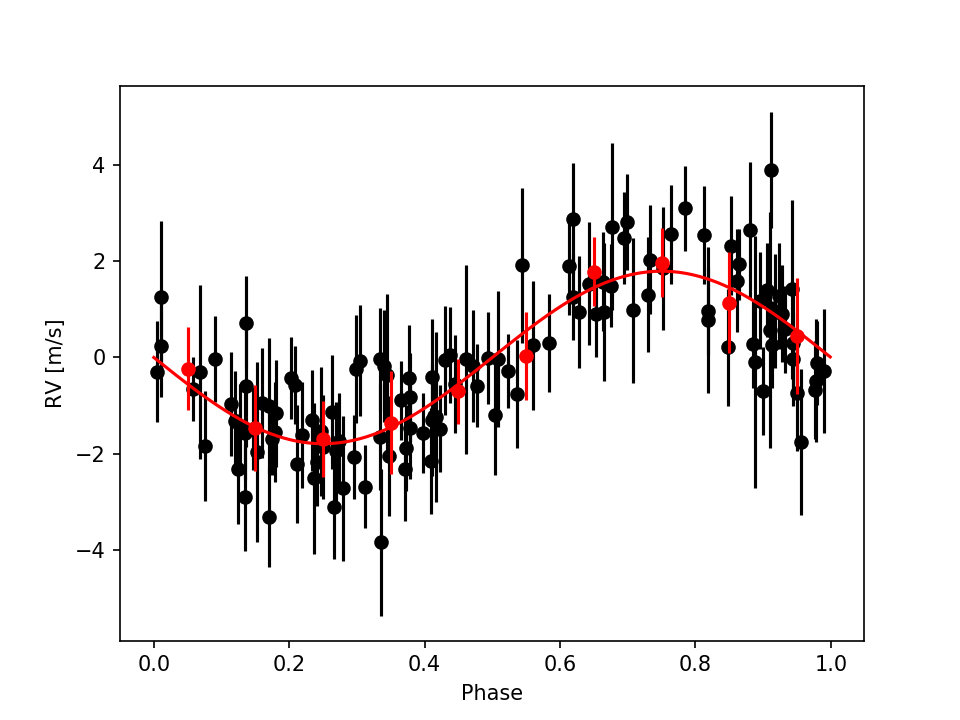}
 \label{fig:appendix_m006:phase_folded}}
 \caption{Planetary candidate in the GJ 21 system. \textit{a)} GLS periodogram  of the RV time series after subtracting the Base Model and the GP activity signal. The vertical blue dashed line mark the orbital period of the candidate, while the vertical red dot-dashed and dotted lines show the stellar rotation period and its first harmonic, respectively. \textit{b)} Phase-folded for the RV curve of the planetary candidate. The red solid line represent the best-fit Keplerian model, while the red circles indicates the binned RVs with the corresponding rms.}
 \label{fig:appendix_m006}
\end{figure*}

\begin{figure*}
   \centering
   \subfloat[][]
 {\includegraphics[width=.45\textwidth]{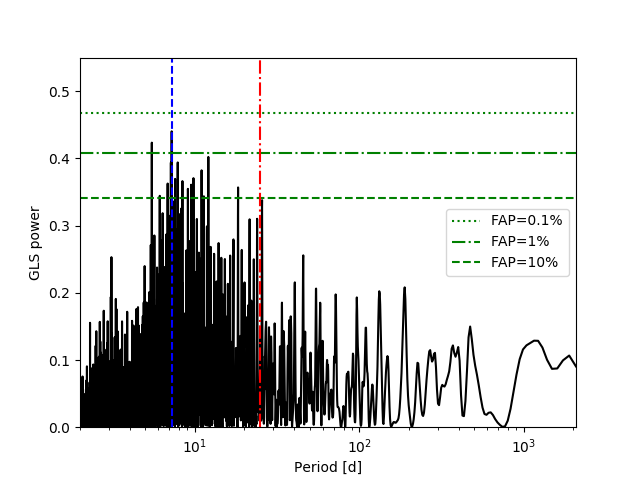}
 \label{fig:appendix_m036:periodogram}}
   \subfloat[][]
 {\includegraphics[width=.45\textwidth]{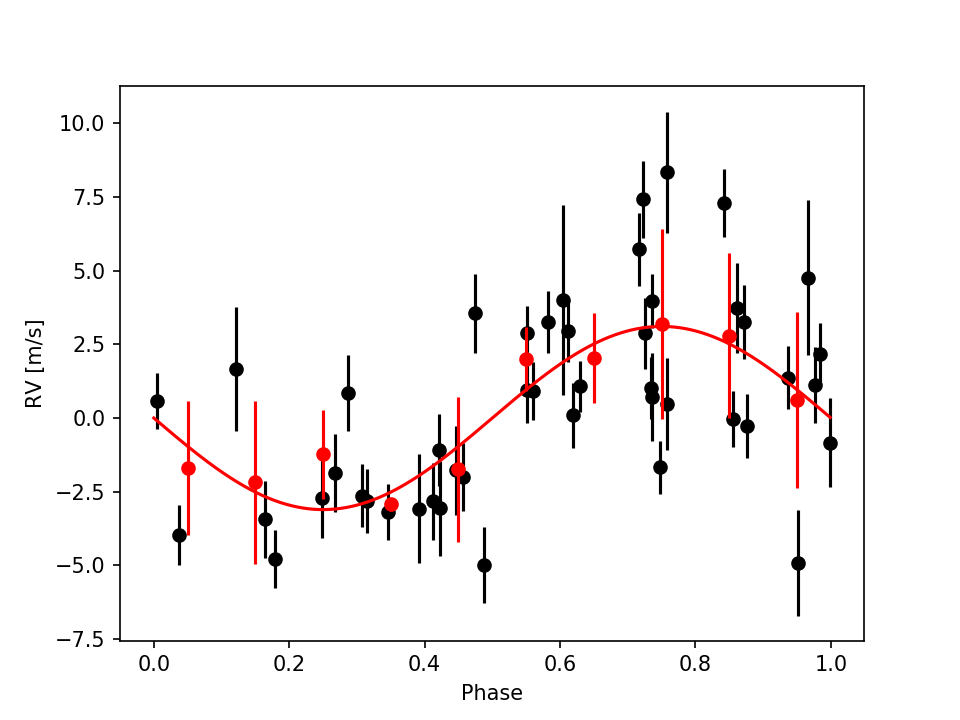}
 \label{fig:appendix_m036:phase_folded}}
 \caption{Planetary candidate in the GJ 1074 system. \textit{a)} GLS periodogram  of the RV time series after subtracting the Base Model. The vertical blue dashed line mark the orbital period of the candidate, while the vertical red dot-dashed line shows the stellar rotation period as derived by \citet{suarezmascarenoetal2017b}. \textit{b)} Phase-folded for the RV curve of the planetary candidate. The red solid line represent the best-fit Keplerian model, while the red circles indicates the binned RVs with the corresponding rms.}
 \label{fig:appendix_m036}
\end{figure*}

\begin{figure*}
   \centering
   \subfloat[][]
 {\includegraphics[width=.45\textwidth]{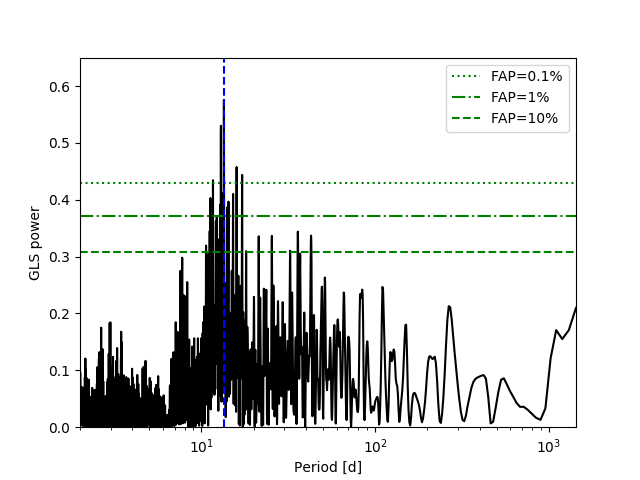}
 \label{fig:appendix_m091:periodogram}}
   \subfloat[][]
 {\includegraphics[width=.45\textwidth]{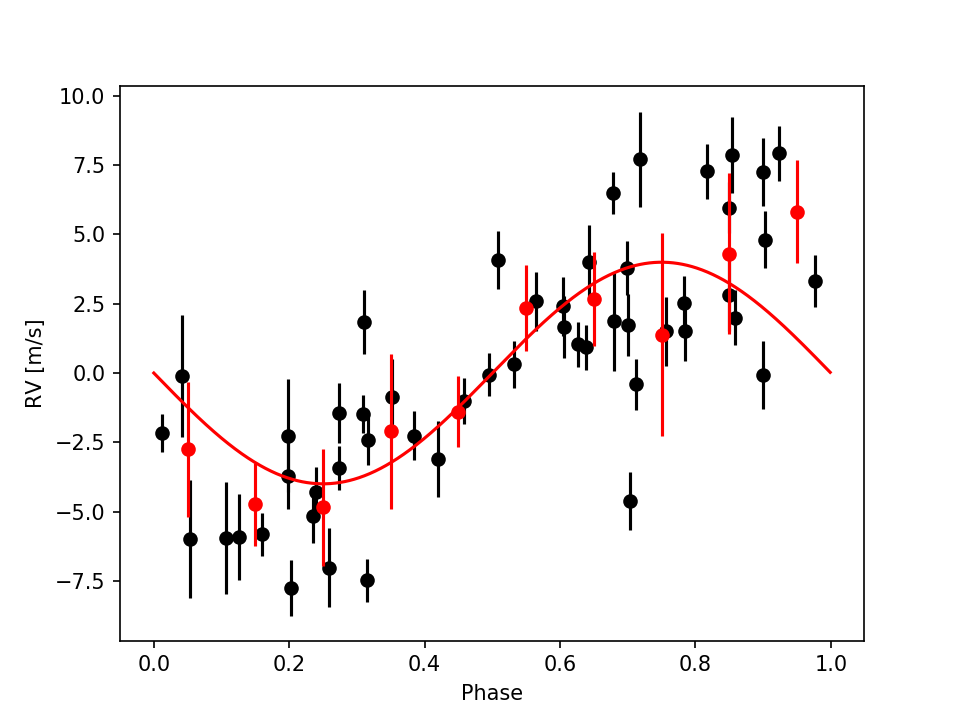}
 \label{fig:appendix_m091:phase_folded}}
 \caption{Planetary candidate in the GJ 9404 system. \textit{a)} GLS periodogram  of the RV time series after subtracting the Base Model. The vertical blue dashed line mark the orbital period of the candidate. \textit{b)} Phase-folded for the RV curve of the planetary candidate. The red solid line represent the best-fit Keplerian model, while the red circles indicates the binned RVs with the corresponding rms.}
 \label{fig:appendix_m091}
\end{figure*}

\begin{figure*}
   \centering
   \subfloat[][]
 {\includegraphics[width=.45\textwidth]{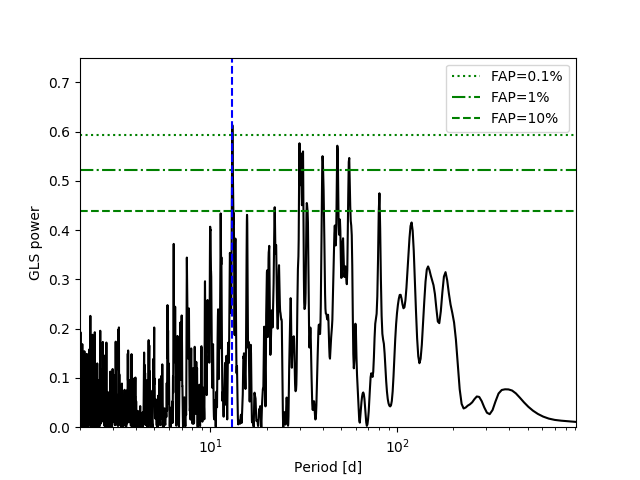}
 \label{fig:appendix_m093:periodogram}}
   \subfloat[][]
 {\includegraphics[width=.45\textwidth]{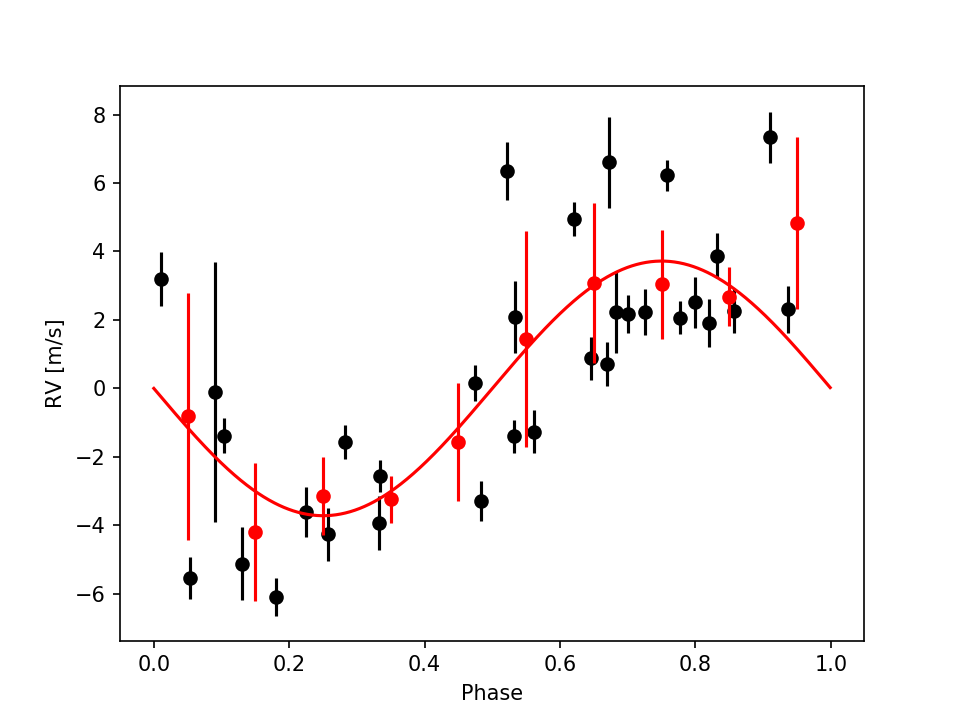}
 \label{fig:appendix_m093:phase_folded}}
 \caption{Planetary candidate in the GJ 548A system. \textit{a)} GLS periodogram  of the RV time series after subtracting the Base Model. The vertical blue dashed line mark the orbital period of the candidate. \textit{b)} Phase-folded for the RV curve of the planetary candidate. The red solid line represent the best-fit Keplerian model, while the red circles indicates the binned RVs with the corresponding rms.}
 \label{fig:appendix_m093}
\end{figure*}

\begin{figure*}
   \centering
   \subfloat[][]
 {\includegraphics[width=.45\textwidth]{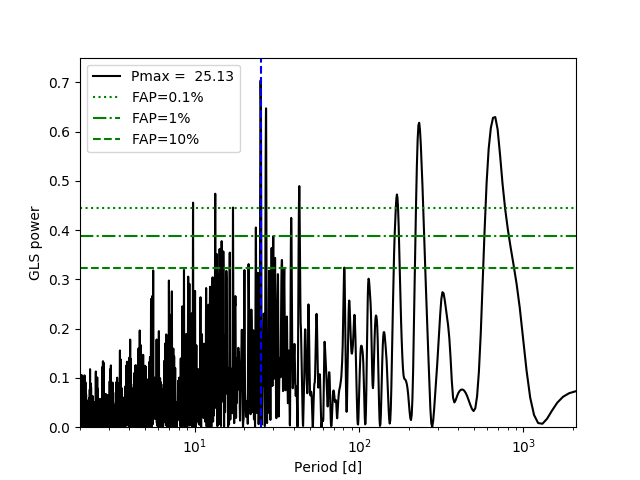}
 \label{fig:appendix_m105:periodogram}}
   \subfloat[][]
 {\includegraphics[width=.45\textwidth]{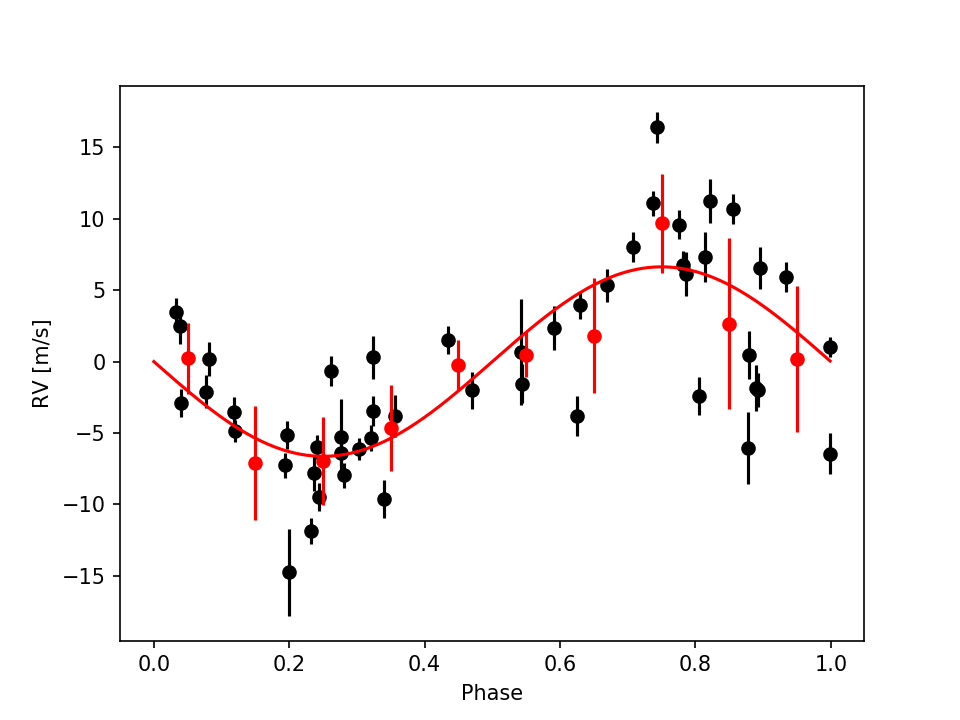}
 \label{fig:appendix_m105:phase_folded}}
 \caption{Planetary candidate in the GJ 3822 system. \textit{a)} GLS periodogram  of the RV time series after subtracting the Base Model. The vertical blue dashed line mark the orbital period of the candidate. \textit{b)} Phase-folded for the RV curve of the planetary candidate. The red solid line represent the best-fit Keplerian model, while the red circles indicates the binned RVs with the corresponding rms.}
 \label{fig:appendix_m105}
\end{figure*}

In Fig. \ref{fig:appendix_m006} to \ref{fig:appendix_m105} the relevant plots describing the planetary candidates detected in this analysis, as discussed in Sect. \ref{sec:candidates}, are shown. The periodic signals in the RV data are shown via the Generalized Lomb Scargle periodogram analysis \citep[GLS,][]{zechkur2009} of the RV time series, after all stellar signals and the Base Model have been removed.

\section{MCMC analysis: best-fit RV models}
\label{app:rv_final_models}

The best-fit models of all the HADES targets fitted in this analysis are summarised in Table \ref{tab:best_fit_models}. The best-fit model components are abbreviated as follow: GP best-fit model includes GP regression as described in Sect. \ref{sec:activity_correction}; BM Base Model from Eq. \ref{eq:baseline_model}; Quad quadratic component added to the Base Model as in Eq. \ref{eq:baseline_model_quad}; Kpl(ecc) full Keplerian component from Eq. \ref{eq:rv_kepl}; Kpl(circ) circular Keplerian component with $e = \omega = 0$ in Eq. \ref{eq:rv_kepl}; Act stellar activity RV signal, modeled as sinewave as discussed in Sect. \ref{sec:activity_correction}..

\clearpage
\onecolumn
\begin{longtable}{lcccc}
\caption{Best-fit RV models of the targets of the survey.}
   \label{tab:best_fit_models}\\
\hline
\hline
Star & Best-fit model & Jitter & Residuals r.m.s. & Activity Correlation   \\
& & $[$m$/$s$]$ & $[$m$/$s$]$ &  \\
\hline
\endfirsthead
\caption{Continued.} \\
\hline
\hline
Star & Best-fit model & Jitter & Residuals r.m.s. & Activity Correlation   \\
& & $[$m$/$s$]$ & $[$m$/$s$]$ &  \\
\hline
\endhead
\hline
\endfoot
\noalign{\smallskip}
GJ 49     & GP + BM + 1Kpl(ecc) & $1.68^{2.81}_{0.34}$ & 1.37 & \\  
\noalign{\smallskip}
GJ 2      & GP + BM & $1.41^{0.24}_{0.23}$ & 1.25 & \\ 
\noalign{\smallskip}
GJ 21     & GP + BM + 1Kpl(circ) & $0.90^{0.36}_{0.41}$ & 0.89 & \\
\noalign{\smallskip}
GJ 119A   & GP + BM + Quad & $1.11^{0.28}_{0.57}$ & 1.21 & \\
\noalign{\smallskip}
GJ 47     & GP + BM & $1.36^{0.22}_{0.20}$ & 1.48 & \\
\noalign{\smallskip}
GJ 119B   & BM & $1.03^{0.67}_{0.50}$ & 1.35 & H$\alpha$  \\ 
\noalign{\smallskip}
GJ 3117A  & BM & $2.27^{0.78}_{0.55}$ & 2.01 & Na~{\sc i}  \\ 
\noalign{\smallskip}
GJ 1030   & BM & $4.40^{1.12}_{0.82}$ & 4.16 & H$\alpha$ \\ 
\noalign{\smallskip}
GJ 4306   & GP + BM & $1.25^{0.16}_{0.16}$ & 1.47 & \\ 
\noalign{\smallskip}
GJ 16     & GP + BM + Quad & $0.93^{0.21}_{0.21}$ & 1.57 & \\
\noalign{\smallskip}
GJ 70     & BM & $2.58^{0.49}_{0.40}$ & 2.49 &   \\
\noalign{\smallskip}
GJ 26     & BM + 2Act & $1.77^{0.28}_{0.24}$ & 1.92 & \\
\noalign{\smallskip}
GJ 15A    & BM + 1Kpl(circ) & $1.46^{0.14}_{0.13}$ & 1.36 & \\
\noalign{\smallskip}
V$^*$BRPsc & BM + Quad & $1.59^{0.23}_{0.20}$ & 1.69 & \\
\noalign{\smallskip}
NLTT 53166 & BM & $2.66^{0.56}_{0.45}$ & 3.30 & \\
\noalign{\smallskip}
GJ 162    & GP + BM & $1.61^{0.31}_{0.28}$ & 1.52 & \\
\noalign{\smallskip}
GJ 150.1B & GP + BM & $1.34^{0.30}_{0.30}$ & 1.70 & \\
\noalign{\smallskip}
GJ 156.1A & GP + BM & $1.21^{0.31}_{0.31}$ & 1.30 & \\
\noalign{\smallskip}
GJ 1074   & BM + 1Kpl(circ) & $2.32^{0.38}_{0.32}$ & 2.73 & \\
\noalign{\smallskip}
GJ 184    & GP + BM & $1.55^{0.29}_{0.30}$ & 1.55 & \\
\noalign{\smallskip}
GJ 272    & BM & $1.95^{0.79}_{0.56}$ & 1.92 & Ca~{\sc ii}, H$\alpha$, He~{\sc i}  \\
\noalign{\smallskip}
GJ 3352   & BM & $2.96^{1.08}_{0.75}$ & 2.99 &   \\
\noalign{\smallskip}
NLTT 21156 & BM & $16.29^{2.12}_{1.75}$ & 15.91 &   \\
\noalign{\smallskip}
GJ 9689   & GP + BM + 1Kpl(ecc) & $1.50^{0.26}_{0.26}$ & 2.64 &   \\
\noalign{\smallskip}
TYC3379-1077-1& BM & $4.36^{1.87}_{1.28}$ & 4.95 &   \\
\noalign{\smallskip}
BPM96441  & BM + Quad & $1.67^{0.64}_{0.50}$ & 2.01 & Na~{\sc i}  \\
\noalign{\smallskip}
TYC2703-706-1& GP + BM & $3.37^{2.47}_{1.84}$ & 2.23 &   \\
\noalign{\smallskip}
StKM1-650 & BM & $5.06^{1.27}_{0.97}$ & 4.17 &   \\
\noalign{\smallskip}
GJ 625    & GP + BM + 1Kpl(ecc) & $1.28^{0.19}_{0.20}$ & 1.65 &   \\
\noalign{\smallskip}
GJ 3942   & GP + BM + 1Kpl(circ) & $2.80^{0.31}_{0.27}$ & 2.64 &   \\
\noalign{\smallskip}
GJ 685    & GP + BM + 1Kpl(circ) & $1.44^{0.35}_{0.32}$ & 1.25 &   \\
\noalign{\smallskip}
GJ 521A   & GP + BM + 1Act & $0.88^{0.24}_{0.26}$ & 0.95 &   \\
\noalign{\smallskip}
GJ 793    & BM & $1.53^{0.34}_{0.27}$ & 1.68 &   \\
\noalign{\smallskip}
GJ 552    & GP + BM & $1.19^{0.28}_{0.35}$ & 1.30 &   \\
\noalign{\smallskip}
GJ 720A   & GP + BM + 1Kpl(ecc) & $1.34^{0.16}_{0.16}$ & 1.39 &   \\
\noalign{\smallskip}
GJ 9440   & BM + 1Act & $2.13^{0.20}_{0.18}$ & 2.35 &   \\
\noalign{\smallskip}
GJ 414B   & BM + 1Act & $1.14^{0.28}_{0.23}$ & 1.46 & Ca~{\sc ii}    \\
\noalign{\smallskip}
GJ 412A   & BM + 1Act & $2.02^{0.19}_{0.16}$ & 2.08 & \\
\noalign{\smallskip}
GJ 694.2  & BM + Quad + 1Act & $2.70^{0.21}_{0.19}$ & 3.27 & Ca~{\sc ii}   \\
\noalign{\smallskip}
GJ 3998   & GP + BM + 2Kpl(circ) & $1.37^{0.30}_{0.19}$ & 1.39 & \\
\noalign{\smallskip}
GJ 408    & BM + 1Act   & $1.65^{0.23}_{0.20}$ & 1.70 &   \\
\noalign{\smallskip}
GJ 450    & BM & $3.36^{0.48}_{0.40}$ & 3.37 &   \\
\noalign{\smallskip}
GJ 9404   & BM + 1Kpl(circ) & $2.95^{0.46}_{0.36}$ & 2.91 &   \\
\noalign{\smallskip}
GJ 606    & BM + 1Act & $2.69^{0.72}_{0.58}$ & 4.50 &   \\
\noalign{\smallskip}
GJ 548A   & BM + 1Kpl(circ)   & $2.91^{0.54}_{0.43}$ & 2.70 & Ca~{\sc ii}   \\
\noalign{\smallskip}
GJ 686    & GP + BM + 1Kpl(circ) & $1.43^{0.29}_{0.28}$ & 1.91 &   \\
\noalign{\smallskip}
GJ 3649   & BM & $1.95^{0.49}_{0.40}$ & 2.21 & H$\alpha$ \\
\noalign{\smallskip}
GJ 4092   & BM & $3.41^{0.46}_{0.39}$ & 3.55 &   \\
\noalign{\smallskip}
GJ 731    & BM + 1Act & $1.01^{0.27}_{0.23}$ & 1.16 &   \\
\noalign{\smallskip}
GJ 2128   & BM & $1.74^{0.42}_{0.34}$ & 1.90 & H$\alpha$ \\
\noalign{\smallskip}
GJ 3997   & GP + BM + Quad & $2.72^{0.32}_{0.29}$ & 4.26 &   \\
\noalign{\smallskip}
GJ 399    & BM + 1Kpl(circ) & $1.91^{0.42}_{0.35}$ & 2.40 &   \\
\noalign{\smallskip}
GJ 740    & GP + BM + 1Kpl(circ) & $1.20^{0.39}_{0.27}$ & 0.92 & Ca~{\sc ii}, H$\alpha$, Na~{\sc i}   \\
\noalign{\smallskip}
GJ 476    & BM & $1.87^{0.54}_{0.41}$ & 1.79 &   \\
\noalign{\smallskip}
GJ 3822   & BM + 1Kpl & $4.24^{0.57}_{0.48}$ & 4.24 &   \\
\noalign{\smallskip}
GJ 4057   & GP + BM & $1.46^{0.21}_{0.19}$ & 1.74 &   \\
\noalign{\smallskip}
\end{longtable}

\twocolumn

\end{appendix}

\end{document}